\documentclass[twocolumn]{aastex631}

\usepackage{natbib}
\usepackage{multirow}
\usepackage{xcolor}

\shorttitle{Star-Planet compositional correlation}
\shortauthors{Plotnykov et al.}

\begin{document}
\defcitealias{Brinkman2024}{B24}
\defcitealias{Adibekyan2021}{A21}
\defcitealias{Adibekyan2024}{A24}
\defcitealias{paperI}{paper I}

\title{Evidence of 1:1 slope between rocky Super-Earths and their host stars}

\author[0000-0002-9479-2744]{Mykhaylo Plotnykov}
\affiliation{Department of Physics, University of Toronto, 27 King's College Cir, Toronto, ON M5S, Canada}

\author[0000-0003-3993-4030]{Diana Valencia}
\affiliation{Department of Physics, University of Toronto, 27 King's College Cir, Toronto, ON M5S, Canada}

\author[0009-0000-2200-131X]{Alejandra Ross}
\affiliation{William H. Miller III Department of Physics and Astronomy,
Johns Hopkins University, 3400 N Charles St, Baltimore, MD 21218, USA}

\author[0000-0001-6533-6179]{Henrique Reggiani}
\affiliation{Gemini South, Gemini Observatory, NSF's NOIRLab, Casilla 603, La Serena, Chile}

\author[0000-0001-5761-6779]{Kevin C. Schlaufman}
\affiliation{William H. Miller III Department of Physics and Astronomy,
Johns Hopkins University, 3400 N Charles St, Baltimore, MD 21218, USA}

\begin{abstract}
    The relationship between the composition of rocky exoplanets and their host stars is fundamental to understanding planetary formation and evolution. However, previous studies have been limited by inconsistent datasets, observational biases and methodological differences. This study investigates the compositional relationship between rocky exoplanets and their host stars, utilizing a self-consistent and homogeneous dataset of 21 exoplanets and their 20 host stars. By applying sophisticated interior structure modeling and comprehensive chemical analysis, we identify a potential 1:1 best-fit line between the iron-mass fraction of planets and their host stars equivalent with a slope of $m = 0.94^{+1.02}_{-1.07}$ and intercept of $c = -0.02^{+0.31}_{-0.29}$. This results are consistent at the 1$\sigma$ level with other homogeneous studies, but not with heterogeneous samples that suggest much steeper best-fit lines. Although, our results remain tentative due to sample size and data uncertainties, the updated dataset significantly reduces the number of super-Mercuries from four to one, but it remains that several high-density planets are beyond what a primordial origin would suggest. {The planets in our sample have a wider range of compositions compared to stellar equivalent values, that could indicate formation pathways away from primordial or be the result of random scattering owing to current mass-radius uncertainties as we recover the observed outliers in mock population analysis $\sim15\%$ of the time. To truly determine whether the origin is primordial with a 1:1 true relation, we find that sample of at least 150 planets is needed and that stars that are iron enrich or depleted are high value targets.}
\end{abstract}


\section{Introduction}
    The growing population of purely rocky exoplanets\footnote{\url{https://exoplanetarchive.ipac.caltech.edu}} has opened a new frontier in exoplanetary sciences, offering an unprecedented opportunity to compare the compositions of these planets to their host stars (e.g. refractory ratio, iron-mass fraction space). 
    Given that the final composition of a planet is the end result of a complex chemical and dynamical formation history, comparing stellar-planet compositions may reveal pathways of formation. 

    {Starting with planetesimals, that are presumed to have homogenous composition (similar to the refractory ratios of the host star \citealt{Larimer1967,Grossman1972}) under efficient radial mixing in the protoplanetary disk, the planet grows through a series of processes that may or may not alter their primordial composition \citep{Asphaug2006,Carter2015,Scora2020,Clement2021}.}
    In the Solar System, there is a moderate variety in major element composition as seen by the iron enrichment of Mercury and stony-iron meteorites as well as the iron-depletion of the Moon and aubrite meteorites, with values in between \citep{Keil1989,Morgan1980,Lodders2003,Wieczorek2006,Margot2018}. 
    Proposed mechanisms that result in major chemical sorting so far include those that are based on differing Fe, Si and Mg properties early in the evolution, such as, fractionation driven by the differences in the condensation temperatures of Fe vs Mg-Si \citep{Lewis1972, Grossman1972, Aguichine2020}; photophoresis, which causes Fe vs Mg-Si dust grains to drift differently due to differences in conductivity and density when heated by stellar radiation \citep{Krauss2005, Wurm2013, Cuello2016}; increased nucleation of Fe vs Mg-Si pebbles due to differences in grain size and surface tension that affect growth under the streaming instability \citep{Johansen2022}; or those that are based on losing mantle material once the planet has differentiated into a silicate mantle over an iron core, such as giant impact collisions \citep{Benz1988,Benz2007,Asphaug2014,Clement2021,Scora2024,Ferich2025} and for small bodies (i.e. Mercury-sized), photoevaporation of the mantle \citep{Ito2021}. 
    Thus, comparing the relative abundances of planetary bodies to that of their star, offers the possibility to understand which mechanisms form planets.
    
    Chemical comparisons had been limited only to the Solar System or a few planets \citep{Santos2015} until recently, when enough data of rocky exoplanets with good precision had grown to several tens of planets. 
    \citet{Plotnykov2020} compared a population of allegedly rocky planets to that of the population of planet-hosting stars and found that planets exhibit a greater diversity in core-mass fraction (CMF), iron to magnesium (Fe/Mg) and iron to silicate (Fe/Si) ratios relative to the stellar population.
    Subsequent studies made one-to-one comparisons in an effort to uncover any possible correlations \citep{Adibekyan2021,Schulze2021,Wang2022,Liu2023,Unterborn2023,Brinkman2024,Adibekyan2024}. 
    Notably, \citet{Adibekyan2021} suggested that super-Mercuries form a distinct population around stars that are enriched in core to mantle material (Fe/Mg+Si) and proposed a strong linear relationship with a slope steeper than a 1:1 line between iron-to-silicate mass fraction of planets and their host stars (slope of $>4$). This result has been revised in \citet{Brinkman2024} and \citet{Adibekyan2024} works such that a super-Mercury subpopulation is no longer present, but the steep compositional slope remains.

    {However, most of these different studies have a shortcoming that stems from considering separately the elemental abundances of stars versus their stellar mass and radius. 
    Given that the latter affect inferences on the planetary mass-radius (M-R) and consequently the planetary interior composition, ensuring consistency between stellar composition, mass and radii is expected to update the planetary characteristics.}
    To address this issue and ensure consistency, two groups \citep{Brinkman2024, paperI} have reanalyzed the stellar data and updated planetary M-R accordingly. 
    In our companion paper, hereafter \citetalias{paperI} \citep{paperI}, we explain in detail the procedure to obtain a consistent stellar data set and perform simple planet-stars comparisons. 

    In this paper, we delve deeper into these comparisons to uncover any relationship by: using a comprehensive interior structure model that thoroughly accounts for uncertainties in major element compositions (Mg, Si, Fe) and using a Bayesian statistical analysis which is more appropriate than other statistical methods used so far \citet{Hogg2010,Andreon2015} as well as properly accounting for the exoplanets at the rocky-to-volatile composition boundary. 
    That is, we weigh each exoplanet proportional to the total probability of them being rocky given the uncertainties, thus including possibly volatile planets in a straightforward manner.
    This method is superior in comparisons to assuming negative CMF (non-physical) and more consistent than having assumptions about the total volatile content, which is a dimensionally degenerate problem. 
    
    With the self-consistent data and our careful methodology we find that, although planets have a larger range in composition than their stars, there is an intriguing 1:1 relation. 
    We investigate these discrepancies between our result and that of \citet{Adibekyan2021}, \citet{Adibekyan2024} and \citet{Brinkman2024} (hereafter \citetalias{Adibekyan2021}, \citetalias{Adibekyan2024} and \citetalias{Brinkman2024}) by applying our methodology to their data and find that the main culprits are: the treatment of stellar data, choice of transit depth (more so than radial velocity) and the statistical modeling, such as consideration for the planets at the volatile rocky boundary.
    While a 1:1 relation is supportive of a primordial origin and thus compelling, the sample size and error bars are too large to make definite conclusions as a zero slope is also consistent with the data at the $1 \sigma$ level. 
    Therefore, we investigate how large of a sample of exoplanets is needed in order to firmly establish or rule out a direct connection between stars and their planets.
    It is important to note though, that although the data set we obtain is self consistent in terms of stellar data, it suffers from being observationally heterogeneous with transit depths and radial velocities being obtained using different instruments and retrieval codes. 
    
\section{Methods}
\subsection{Data selection}
\label{data_selection}
 
    In our previous study \citet{Plotnykov2020} we had chosen planets within the rocky M-R region (see Fig. \ref{fig:femf}) with good precision ($\Delta R/R, \Delta M/M <25\%$). 
    For this study we select exoplanets among that sample that orbit FGK stars with high-resolution optical spectra (R$>30000$) and have high signal-to-noise ratios (S/N$>50$ at $ \lambda \approx 6000$ \AA) that enables host star chemical characterization with good precision.
    We also select for the most likely rocky exoplanets by checking that the equilibrium temperature is beyond the supercritical point of water ($T=650K$) such that any volatile layer would be in gaseous form, vulnerable to atmospheric loss under high irradiation. 
    We consider that their total mass and radius is consistent with a purely rocky composition. 
    However, we acknowledge the possibility that some of these exoplanets may have a thick high molecular weight atmosphere, either CO/CO$_2$ ('puffy Venuses' applicable to $T_{eq}> 1500$K, \citet{Peng2024}) or water (vapor planets applicable to $T_{eq}>650$K, \citet{Dorn2021}) above a magma ocean so that their true rocky interior would be denser. 
    Excluding these scenarios means our inferred Fe-MF is a minimum value. 
    That is, planets can always have a larger proportion of iron to offset a volatile layer. 
    
    Our sample consists of 21 exoplanets and 20 host stars, for which we derive homogeneous and physically self-consistent host star data of effective temperature, surface gravity, metallicity, masses and radii ($T_{eff}, log(g), [Fe/H], M_{st}, R_{st}$) (presented in \citetalias{paperI}). 
    Our methodology is described in \citet{Reggiani2022, Reggiani2024} and summarized here. 
    We use the classical spectroscopy analysis to derive the initial set of photospheric stellar parameters (${T_{eff}, log(g), [Fe/H]}$), which we than combine with GAIA astrometry and high-quality photometry data in our isochrone analysis (see \citetalias{paperI} for more details).
    Both spectroscopic and photometric data is coupled together to derive the stellar parameters {$T_{eff}, log(g), [Fe/H], M_{st}, R_{st}$} using a Markov Chain Monte Carlo (MCMC) framework.
    We iterate the fitting process multiple times to achieve self-consistency across all stellar parameters.
    This ensures self-consistency and robustness, reducing degeneracies and biases that could otherwise affect the accuracy of the abundances.
    Subsequently, we obtain individual elemental abundances ([Fe/H], [Mg/H], [Si/H], [Ca/H], [Al/H], [Ni/H]) directly from spectroscopy data using equivalent width-based calculation and manually verifying each absorption line feature, while excluding any fits that failed due to blending, line saturation or some other reason (see \citetalias{paperI}).
    
    For comparison, \citetalias{Adibekyan2021}/\citetalias{Adibekyan2024} used a similar method to derive their abundances, but lacked the isochrone analysis that updates $M_{st}, R_{st}$ and ensures consistency. 
    On the other hand, \citetalias{Brinkman2024} uses a data-driven machine learning model (\texttt{KeckSpec}, a version of \texttt{Cannon} \citet{Ness2015}) that uses \citet{Brewer2016,Brewer2018} data as the training set. 
    Although convenient and fast, with this method it is difficult to assign individual uncertainties to predictions.     
    We obtain similar results in terms of $M_{st}, R_{st}, T_{eff}, log(g)$, but our physically self-consistent approach yields different elemental abundances with important consequences.
    We attribute this discrepancies to the different algorithms and selection processes employed. 
    In addition, in contrast to all other studies, we apply non-local thermal equilibrium (non-LTE) corrections to all the elemental abundances, thereby including an important source of uncertainty. 
    These corrections affect mostly the [Fe/H] and [Si/H] values, but not the [Mg/H].

    With self-consistent stellar data, we use radial velocity and transit measurements found in the NASA Exoplanet Archive \citep{Akeson2013} to update planetary M-R of all planets. 
    Of importance is that most of the exoplanets in our sample have been observed and analyzed by multiple groups and techniques (20 out of 21). 
    We select those that have the largest number of measurements and highest precision, while making sure that transit observation comes from the same source instrument. 
    Our choice for RV and transit data differs in some cases from those of \citetalias{Adibekyan2021}, \citetalias{Adibekyan2024} and \citetalias{Brinkman2024} yielding another source for discrepancy in planetary M-R data (see Sec. \ref{sec:comparison} for more details).

\subsection{Interior structure model}
\label{rocky_model}
    To infer the planetary composition given planetary M-R, we use the interior structure model \texttt{SuperEarth} developed by \citealt{Valencia2006,Valencia2007} and revamped by \citet{Plotnykov2020}.
    \texttt{SuperEarth} code is a numerical interior integrator that solves for the mass, density, pressure, gravity and temperature as a function of radius, given a total mass, core-mass fraction (CMF) and chemical mineralogy ($\chi$).  
    That is, \texttt{SuperEarth} is a forward model that calculates the planet's radius $R$ given its mass $M$ and interior structure: $\mathrm{R=R(M; CMF,\chi)}$. 
    In our analysis, we consider mineralogies based on the Earth and account for the following variations:
    \begin{enumerate}
        \item Variable degree of core differentiation, characterized by the iron content in the mantle:
        \[
        \mathrm{xFe} =1-\mathrm{Mg}\#=1-\frac{\mathrm{Mg}}{\mathrm{Mg}+\mathrm{Fe}}
        \]
        \item Variable silicon content in the core (xSi), which impacts the Fe/Si ratio as well as the total amount of light material in the planet. 
        \item Variable Mg/Si ratio controlled by the proportions of mantle minerals (e.g. bridgmanite vs wustite, $\mathrm{(Mg,Fe)SiO_3}$ vs $(Mg,Fe)O$).
    \end{enumerate}

    To estimate interior composition from M-R data, we combine our forward interior model with an affine invariant Markov Chain Monte Carlo (MCMC) sampler \texttt{EMCEE} \citep{Foreman2013}, assuming that the M-R constraints are independent and follow a Gaussian distribution in our log-likelihood function. 
    We check the effects of existing correlations between the stellar M-R posterior results (see \citetalias{paperI}), but find that they do not translate to the planetary parameters, as the latter is dominated by uncertainties in the radial velocity/transit signal. 
    
    For our MCMC setup, we follow the computational procedure outlined in \citet{Plotnykov2020} and assume uniform distributions within physical bounds for all the parameters (M, R, CMF and $\chi$), but for $\chi$ we specifically assume that $\mathrm{xSi}=\mathcal{U}(0,20)\%$ by mol and $\mathrm{xFe}=\mathcal{U}(0,20)\%$ by mol, while for the amount of wustite [(Mg,Fe)O], which affects the Mg/Si ratio, we adopt a fixed value (xWu$=20\%$). 
    We have increased the xSi range compared to our previous study \citep{Plotnykov2020} to encompass the wide range of proposed compositions consistent with Earth's density profile \citep{Fischer2015, Zhang2018, Umemoto2020, Hikosaka2022}. 
    Meanwhile, fixing xWu has minimal impact on the density profile of the planet and consequently on CMF ($\Delta\mathrm{CMF}<0.01$ \citep{Plotnykov2024}). 

    We had previously found that CMF values are highly dependent on prior assumptions (e.g. xSi, xFe)\footnote{The change in CMF can be up to $\Delta \mathrm{CMF} \sim 0.07$.}, while to the iron-mass fraction (Fe-MF) of planets remains highly invariant to most mineralogy priors and where variations do occur, they tend to be minor \citet{Plotnykov2024}. 
    Thus, for this study we focus on Fe-MF values and obtain the full range by considering, a-posteriori, a mineralogical model consistent with the density profile that fits the planetary M-R. 
    For this, we consider minor minerals such as corundum ($\mathrm{Al_2O_3}$) and calcium perovskite ($\mathrm{CaSiO_3}$) in the mantle because these are major carriers of Al and Ca (xAl, xCa) as well as we vary nickle in the core for the same reason (xNi). 
    In contrast to \citetalias{paperI}, we also vary the Mg/Si ratio controlled by the proportions of mantle minerals (e.g. bridgmanite vs wustite). 
    Note that in the case when the mantle Mg/Si molar ratio is less than 1, in the absence of experiments, we assume that there is no wustite in the mantle (xWu=0) and the remaining silica is incorporated into quartz ($\mathrm{SiO_2}$, xQtz). 
    However, experiments should be carried out to better understand the mineralogy and equation of state (EOS) at these low Mg/Si values, as there is a considerable number of stars that display them ($50\%$ of APOGEE stars have Mg/Si$<0.7$ by weight \citet{APOGEE}).

    Thus, owing to the lack of geophysical and theoretical constraints in these relative abundances, we adopt host star elemental ratios by weight for Mg/Si, Ca/Mg, Al/Mg and Ni/Fe as priors to constrain the added mineralogy. 
    Almost all of these ratios are consistent among stars with different metallicities. 
    The exception is the Mg/Si ratio, which has the most variation among the different stars when compared to GALAH \citep{GALAH}, APOGEE \citep{APOGEE}, HYPATIA \citep{HYPATIA} surveys as well as the variation within the solar system \citep{Anders1964}. 

    Adding these minor minerals a-posteriori has a negligible effect in the density and CMF, given that their EOS's are fairly similar to the major chemical phases our interior structure model considers\footnote{The difference between using the minor mineralogies for calculating the structure versus after is a difference in radius of $\sim 1$km for a 1 earth-mass planet.}.
    However, they are needed to obtain the true Fe-MF range allowed for a planet consistent with its M-R. 
    For example, depending on the amount of these minor elements, the Fe-MF can change by $\sim 0.05$ \citep{Plotnykov2024}. 
    
    Having this chemical treatment allows us to report realistic Fe-MF's as well as refractory ratios (Fe/Si, Fe/Mg, etc.) for all planets. 
    We report the results in terms of the median ($50^{\mathrm{th}}$ percentile) and 1-standard deviation ($1 \sigma$ confidence interval, $16^{\mathrm{th}}$ and $84^{\mathrm{th}}$ percentile) values from the derived distributions.

\subsection{Stellar equivalent chemical model}
\label{chem_model}
    To directly compare planets to their host stars, we employ our chemical interior model that translates stellar abundances ([Fe/H], [Mg/H], [Si/H], [Ca/H], [Al/H], [Ni/H]) into planetary equivalent interior structure (Fe-MF$_{st}$).
    Hence, we assume the same minerals present in our planetary structure model and consider the same range in the level of differentiation (xFe) and amount of alloy in the core (xSi), resulting in a $\mathrm{CMF}>0$.\footnote{If all of the iron were to reside in the mantle ($\mathrm{CMF}=0$), the Fe-MF$_\mathrm{st}$ would decrease by $2-3$ points compared to the differentiated case}.
    
    Given this chemical model, we equate the stellar refractory ratios of Fe/Mg, Fe/Si, Mg/Si, Ca/Mg, Al/Mg, Ni/Fe to those of a planetary interior and obtain the CMF$_\mathrm{st}$ and Fe-MF$_\mathrm{st}$ such a planet would have. 
    Note that in this chemical model, both CMF$_\mathrm{st}$ and Fe-MF$_\mathrm{st}$ are independent of total planetary M-R.
    We express this model as follows: Fe-MF$_{st} = Z(\mathrm{CMF},\chi) = Z\mathrm{(CMF, xWu (xQtz), xFe, xSi, xNi, xCa, xAl)}$.
    We use a Monte Carlo simulation to retrieve the Fe-MF$_{st}$ posterior and sample the interior parameters 50,000 times given host star refractory ratios.
        
\subsection{Statistical method}   
    \label{correlation}   
    Owing to the small sample size (N=21), we choose to investigate possible correlations between stellar and planetary parameters with a simple linear function ($Y=mX+c$) as to avoid over-fitting the data.
    To properly evaluate the best-fit line, we use \texttt{STAN} codebase \citep{stan} and set-up a Hamiltonian Monte Carlo (HMC) with a No-U-Turn sampler (NUTS) to find the posterior for $m$ (gradient) and $c$ (intercept) values.
    This method is superior in comparison to ordinary least-square (OLS) or orthogonal linear regression (ODR), due to a multitude of factors \citet{Hogg2010,Andreon2015}.
    Specifically, in the context of our sample, a Bayesian method mitigates the effect of outliers and allows for fitted parameters to have non-Gaussian posterior distributions ($m$ and $c$), both of which are extremely important factors owing to the small sample size (see Sec. \ref{sec:comparison} for comparison). 
    More importantly, however, is that a Bayesian analysis has greater flexibility on the assumed probability density of the input data, in contrast to both OLS and ODR treatments that assume symmetrical errors (Gaussian distribution). 
    We find that most stellar and planetary posterior parameters (e.g. Fe-MF$_\mathrm{st}$, Fe-MF$_\mathrm{pl}$) have skewness in their distributions, thereby we assume each parameter follows a skewed-Gaussian ($\mathcal{SN}$) distribution in our likelihood ($\mathcal{L}$) analysis.

    The probability density function for the skewed-Gaussian is defined as:

    \begin{equation}
        \mathcal{SN}(x|\xi,\omega,\alpha) \sim \phi\left(\frac{x - \xi}{\omega}\right) \Phi\left(\alpha \left(\frac{x - \xi}{\omega}\right)\right),
    \end{equation}
    
    where $\alpha$ is the shape parameter, $\omega$ is the scale parameter and $\xi$ is the location parameter that describe the skewed-Gaussian, while $\phi$ and $\Phi$ are the standard normal probability and cumulative density functions respectively.
    Thus, the joined likelihood function for the linear fit of stellar and exoplanet variables ($X$, $Y$) is defined as the product of skewed-Gaussian distributions:

    {
    \begin{equation}
        \mathcal{L} \sim \prod_{i=1}^{n} \mathcal{SN}(x_i|\xi_{x;i},\omega_{x;i},\alpha_{x;i}) \mathcal{SN}(y_i|\xi_{y;i},\omega_{y;i},\alpha_{y;i}),
    \end{equation}
     where $x_i$ and $y_i$ are the i-th dataset pair of $X$ and $Y$, assuming that $y_i=x_i \, m+c$.
     }

    This linear best-fit yields an empirical relationship that may be interpreted in the context of formation scenarios.
    However, a different method to determine whether the planet-star data supports different formation scenarios is to use a Hierarchical Bayesian Modeling (HBM) that ties all the data at a population level. 
    After running some tests to obtain hyperparameters that could describe a systematic shift from or an intrinsic scatter around stellar composition due to formation reprocessing, we deemed the sample is too small and the errors are too large at the moment. 
    This is worth revisiting once enough data is acquired (for example after PLATO launches \citet{PLATO}).
    
    In addition to a sound statistical treatment, we need to pay careful attention to planets that straddle the volatile-rocky boundary (e.g. rocky threshold radius--RTR or largest size for an MgSiO$_3$+MgO planet) as well as the pure iron line. 
    These exoplanets have part of their probably density functions in M-R space within the rocky region and part of it outside.
    One way to model them is to restrict the composition solutions only to the rocky subset (employed by \citet{Plotnykov2020} and \citet{Adibekyan2021}), but this artificially increases the certainty in their composition leading to small error bars that would anchor any inferred relationship slope. 
    Without the knowledge on the absence of an envelope, this method may yield misleading results. 
    Another way is to allow planets to have negative CMFs to represent the M-R solutions that call for volatiles \citep{Brinkman2024}. 
    This is done by extrapolating fits to $\mathrm{CMF}=\mathrm{CMF}(M,R)$ that are obtained within the rocky region, to negative values. 
    This, of course, is non-physical and highly dependent on the extrapolation scheme.
    
    Instead here, we propose a simple statistical approach to include these exoplanets by weighting them (with weights $w_i$) in the linear fitting according to the area of the probability density, given the M-R uncertainties, that falls within the rocky region $w_i=P_\mathrm{rocky}$. 
    That is, exoplanets that have 3 $\sigma$ confidence of being within the rocky region will have $w_i\approx1$.    
    {We find that removing exoplanets that are near to the RTR boundary does not alter our weighted best-fit solutions, since these planets have $w_i<0.5$.}
    This physical-statistical approach allows us to adequately and swiftly include exoplanets with error bars at the boundaries of the rocky region and determine the goodness-of-fit of our linear model.
    
    Lastly, to test for robustness, especially given the small sample size, we perform a leave-one-out cross-validation (LOOCV) test, where we re-fit our dataset by removing one data point at a time, identifying whether or not the observed linear fits are dependent on a few key exoplanet-star pairs. 
    This validates whether the sub-sample (N-1) supports or refutes the presence of a linear relationship and exposes the planets that affect any inferences the most.

\section{Results}
\subsection{Link between planets and stars}
    To compare compositions of exoplanets to their host stars (planet-star) we apply our most sophisticated planetary interior model with full mineralogy consideration and our chemical model that translates stars refractory abundances to planet equivalent parameters (see Sec. \ref{data_selection}).
    We search for correlations in various different parameters and focus on the refractory ratios (Fe/Si, Fe/Mg, etc.), Fe-MF, age of the system and normalized density with the goal of identifying any links or compositional trends. 
    Note that our fiducial dataset assumes non-local thermal equilibrium (non-LTE) correction for stellar abundances and hence Fe-MF$_{st}$.
    
    We present the resulting planet-star compositions in Table \ref{tab:planet}, \ref{tab:star} and include for comparison planetary CMF results that are not physically self-consistent (see Fig. \ref{fig:data_full}).
    Given the updated radius of 55~Cnc~e, this planet is no longer considered rocky as it is $\sim 3 \sigma$ above the rocky-threshold-radius (RTR; $\rho/rho_\oplus<0.8$) and may have a significant amount of volatiles, making it impossible to constrain the rocky interior.
    Additionally, we identify two more planets that seem to have volatiles (more on this in \citetalias{paperI}), WASP-47~e and Kepler-36~b, given the fact that they straddle the RTR boundary. 
    As explained, we weight the planets contributions to the fits according to how much their M-R data intersects the rocky versus volatile/over-dense regions.

\subsubsection{Relationship in chemical ratios}
\begin{figure*}
        \centering
        \includegraphics[width=0.9\textwidth]{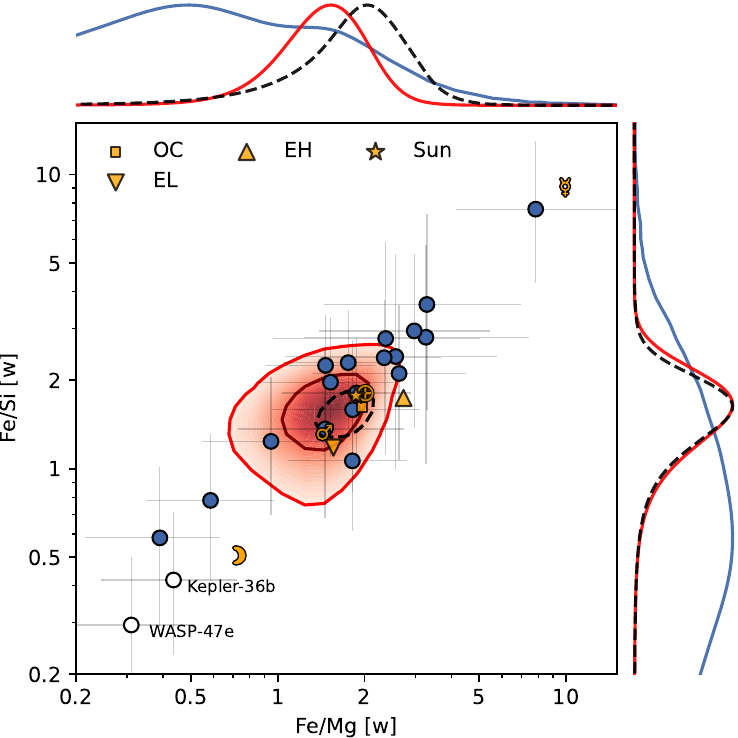}
        \caption{Refractory ratio of stars compared to their exoplanets as populations. The host stars population is shown in red at contours of 1 and 2 $\sigma$ values, while the dotted contour corresponds to the GALAH stellar population. The marginal distribution of Fe/Mg and Fe/Si are shown outside of the scatter plot for stars (red), all exoplanets (blue) and GALAH (dotted) populations. Exoplanets above the RTR, which require volatiles, are indicated as white circles and solar system objects are shown in yellow. Notice the axes are in logarithmic scale.}
        \label{fig:ratio}
    \end{figure*}

    We first compare the updated planetary data in a population sense and confirm our observation in \citet{Plotnykov2020} that exoplanets span a wider refractory ratio space (Fe/Mg, Fe/Si by weight) compared to their host stars (see Fig. \ref{fig:ratio}).
    However, unlike the \citet{Plotnykov2020} sample, this updated data, derived from a homogeneous and physically self-consistent stellar treatment consists of a higher concentration of exoplanets around an Earth-like composition, in better agreement with a stellar composition (see Fig. \ref{fig:mr_all}).

\begin{figure}
        \centering
        \includegraphics[width=\linewidth]{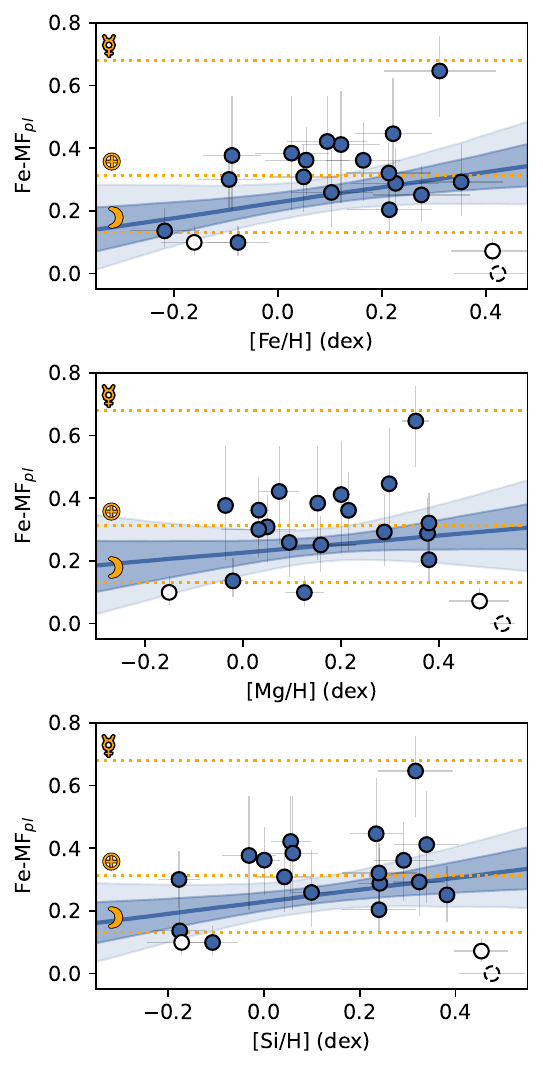}
        \caption{Exoplanet's iron-mass fraction (Fe-MF$_{pl}$) compared to the stellar [Fe/H] (top panel), [Mg/H] (middle panel) and [Si/H] (bottom panel) abundances. The results of our linear best-fit (assuming weighted sample) are shown as blue contours at 1 and 2 $\sigma$ confidence values. Similarly, exoplanets that are above RTR are indicated as white circles, while 55-Cnc~e is a dotted white circle.} 
        \label{fig:femfvsstar}
    \end{figure}

    We directly compare the stellar abundances of the main refractory elements ([Fe/H], [Mg/H], [Si/H]) to the exoplanet iron content (Fe-MF$_{pl}$) using the weighted best-fit analysis and show the results in Figure \ref{fig:femfvsstar}. 
    There appears to be a positive slope between the iron mass fraction of the planet Fe-MF$_{pl}$ and the metallicity of the star as measured by [Fe/H], with a slope of $m=0.24 \pm 0.16$ and intercept of $c=0.23 \pm 0.03$.
    Meanwhile, Fe-MF$_{pl}$ relationships with [Mg/H] and [Si/H] are weaker with shallower slopes ($m<0.2$, see Table \ref{tab:coef_ext}).

    These results are in general agreement with the findings in \citetalias{paperI} (within $1\sigma$), but our improved statistical treatment reveals evidence for a somewhat steeper slope.
    Although, at this point there is no physical interpretation framework for these relationships and should be treated as a test in confirming planet-star link existence.
    These trends continue to be present when comparing the weighted dataset in refractory ratios (Fe/Mg, Fe/Si, etc), indicating a direct correlation between exoplanets and their stars. 
    We find that both Fe/Mg and Fe+Ni/Mg+Si ratios by weight have a slope close to 1:1 ($m\sim0.7$), while the slope is weaker in Fe/Si space ($m\sim0.3$).
    However, these fits do not tell the full story due to the non-linear nature of these ratios, where under-dense planets (CMF$<0.1$) have several times smaller uncertainty compared to Earth-like or over-dense planets.
    For example, we find that the error in Fe/(Mg+Si) ratio is 5 times larger for an Earth-like compared to an iron-depleted planet with the same M-R percentage uncertainties (e.g. HD~136352~b vs K2-265~b).
    Therefore, we cannot use these ratios reliably to evaluate the compositional link between planets and stars; rather, other linear parameters such as Fe-MF$_{pl}$, uncompressed density or normalized density are better options.

\subsubsection{Relationship in iron-mass fraction}
\begin{figure*}
        \centering
        \includegraphics[width=\linewidth]{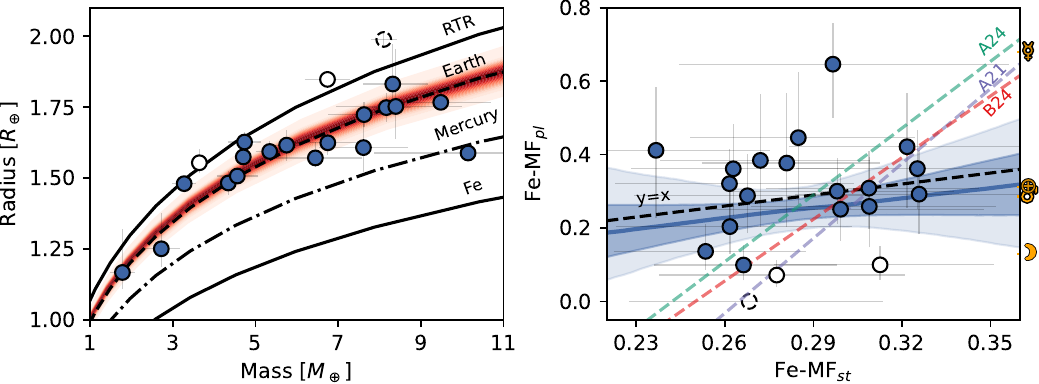}
        \caption{Comparison planets' composition to their stars. Left: Planetary masses and radii in our sample with host stars planet equivalent values as a population shown in red. Right: Direct comparison of iron-mass fraction of stars (Fe-MF$_{st}$) and iron-mass fraction of planets (Fe-MF$_{pl}$), including our linear best-fit (assuming weighted sample) shown as blue contours at 1 and 2 $\sigma$ confidence values. The dashed lines are the ODR best-fit results provided by studies \citetalias{Adibekyan2021} (purple), \citetalias{Adibekyan2024} (green) and \citetalias{Brinkman2024} (orange). Our data and methodology is consistent with a 1:1 compositional relation between planets and stars (Fe-MF$_{pl}$=Fe-MF$_{st}$) shown as a dashed black line, we summarize our best-fit solutions in table \ref{tab:coef_ext}. We use the same color scheme for each star-planet pair as before.}
        \label{fig:femf}
    \end{figure*}

    The comparison in Fe-MF space is the most robust and direct metric for testing the composition link between planets and their host stars.
    With our weighted dataset, we find an intriguing possible 1:1 slope in Fe-MF space, given a slope of $m = 0.94^{+1.02}_{-1.07}$ and an intercept of $c = -0.02^{+0.31}_{-0.29}$, see Figure \ref{fig:femf}. 
    Furthermore, we do not find strong evidence for any sub-populations, either in iron-rich or iron-poor exoplanets across the sampled host stars in agreement with \citetalias{Brinkman2024} and \citetalias{Adibekyan2024}.     
    Notably, treating each exoplanet with equal weights shifts the slope away from a 1:1 relationship (see Sec. \ref{robustness}).
    Thus, we find the choice of statistical methods to be highly important and without proper outlier treatment, the constraints on this relationship may be biased towards steeper/shallower best-fit lines.
    
    Altogether for the population, our data and analysis hints at similar refractory composition between stars and their planets with a 1:1 relationship, but given the large errors in planetary masses and radii, this result is still inconclusive as a zero slope ($m=0$) is also allowed within a $1 \sigma$ confidence interval. 
    This stands in contrast to the steep slope obtained by \citetalias{Adibekyan2021}, \citetalias{Adibekyan2024} and \citetalias{Brinkman2024} using ODR, with slopes that have $m>5$.
    We discuss the multiple reasons for these discrepancies between the studies in section \ref{sec:comparison}.

\subsubsection{Relationship in normalized density}
    We expand our analysis to include other intrinsic properties and possible correlations in our dataset.
    These properties do not involve any interior or chemical structure modeling, therefore we assume that all the data-points are equally weighted. 
    We find most intrinsic parameters, such as mass, radius, effective temperature, period, etcetera to have no significant linear relationships, except for normalized density ($\rho/\rho_\oplus$), as expected, as it tracks the composition of a planet. 
    But unlike composition, that requires interior modeling, normalized density can be calculated in a straight forward manner, $\rho/\rho_\oplus \approx M_{pl}^{0.86}/R_{pl}^3$ where $M_{pl}$ and $R_{pl}$ are in Earth mass and radius units.
    
    \begin{figure}
        \centering
        \includegraphics[width=\linewidth]{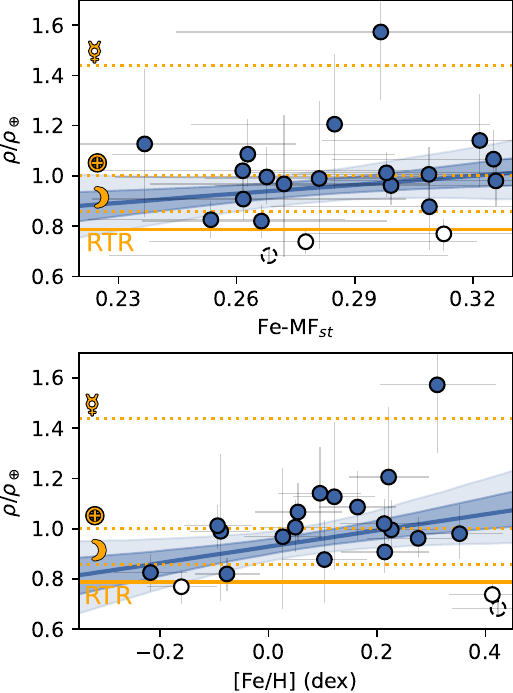}
        \caption{Normalized density as a function of stellar iron-mass fraction (Fe-MF$_{st}$,  top panel) or stellar metallicity ([Fe/H], bottom panel). We use the same color scheme for each star-planet pair as before and show the best-fit results as blue contours at 1 and 2 $\sigma$ confidence values. For these fits we exclude planets that are not presumably rocky  (i.e. unweighted sample of planets with $\rho/\rho_\oplus>0.8$).}
        \label{fig:rho_earth}
    \end{figure}

    Notably, $\rho/\rho_\oplus$ shows a strong positive slope with stellar Fe-MF$_{st}$, with or without exoplanets that are above RTR (see Fig. \ref{fig:rho_earth}).
    Thus, it appears as though $\rho/\rho_\oplus$ may substitute Fe-MF$_{pl}$ for rocky planets, but caution has to be taken for planets that may have volatiles ($\rho/\rho_\oplus<0.8$).
    This is because normalized density changes with total mass even if exoplanet composition is fixed, for example $\rho/\rho_\oplus$ for a Mercury-like exoplanet will change from $1.4$ to $1.5$ given $1M_\oplus$ and $10M_\oplus$ mass exoplanet.
    Although, for exoplanets that are bellow the threshold $\rho/\rho_\oplus<0.8$, the relative changes are higher and it is impossible to accurately determine the interior composition from normalized density (M-R) due to compositional degeneracy.
    Meanwhile, $\rho/\rho_\oplus$ relationship with other stellar parameters is less clear.

\subsection{Best-fit robustness}
\label{robustness}
    For all the best-fits we perform a LOOCV test to verify robustness and identify outliers.
    When using weighted data to constraint Fe-MF$_{pl}$ to stellar abundances ([Fe/H], [Mg/H], [Si/H]) or Fe-MF$_{st}$, we find the relationships to be robust without any major outliers.
    Meaning that removing any exoplanet from the sample does not change the best-fit slope by more than $1 \sigma$ ($\Delta m < 1 \sigma$).
    However, there are a few exoplanets that stand out and should be investigated (HD~136352~b, Kepler-20~b, K2-291~b, Kepler-10~b).
    For example, removing Kepler-10~b weakens the Fe-MF$_{st}$ vs Fe-MF$_{pl}$ best-fit slope the most ($m = 0.4\pm 1$). 
    This is because the host star has an Fe-deficient equivalent composition (Fe-MF$_{st} = 0.266 \pm 3.1$), while the planet is also Fe-poor (Fe-MF$_{pl} = 0.099^{+0.054}_{-0.042}$). 
    On the other hand, removing the only super-Mercury remaining in our sample Kepler-107~c does not change the slope for any of these relationships (with exception of bulk density).
    
    Meanwhile, when equally weighing all exoplanets, the slope increases to $m = 1.89^{+0.78}_{-1.01}$, but excluding the straddling rocky-volatile planets gives a similar to 1:1 $m = 1.29^{+0.77}_{-0.81}$.
    We find that removing WASP-47~e changes the slope back to 1:1 ($m = 1.01^{+1.08}_{-1.45}$), but with larger errors. 
    This exoplanet, we argue is most likely volatile and thus, an outlier to the rocky sample, given that it has a low density around an host star with extremely high [Fe/H], [Mg/H] and [Si/H] abundances (3x that of solar). 
    Thus, without the knowledge of the absence of an atmosphere, assuming planets at the boundary of the rocky region are rocky can lead to incorrect results. 

\subsection{Methodology comparisons}
\label{sec:comparison}
    Using a comprehensive interior structure and sound statistical modeling on a homogeneously and physically self-consistent exoplanet-stellar data, we obtain a planet-star compositional link that is close to 1:1.
    These results stand in contrast to the steep relationships obtained by \citetalias{Adibekyan2021}, \citetalias{Adibekyan2024} and \citetalias{Brinkman2024}. 
    The reason for these discrepant results is due to differences in: 1) retrieved stellar abundances, 2) exoplanet M-R data arising from differences in observed stellar M-R, 3) transit data selection (more so than radial velocity data) and/or 4) the statistical and outlier consideration treatment. 

\begin{figure}
        \centering
        \includegraphics[width=\linewidth]{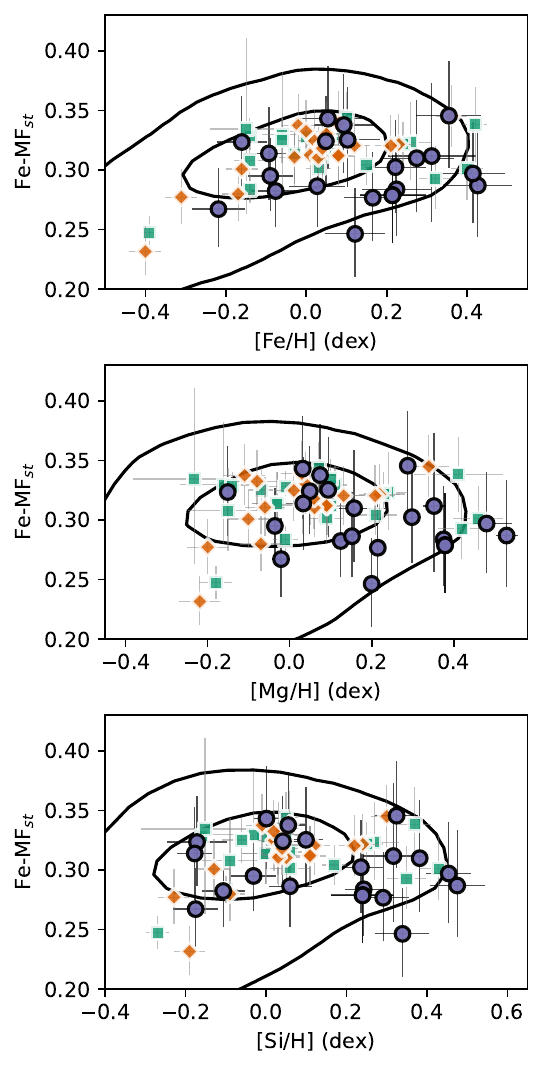}
        \caption{Host star iron-mass fraction equivalent (Fe-MF$_{st}$) as a function of stellar refractory abundances [Fe/H] (top panel), [Mg/H] (middle panel) and [Si/H] (bottom panel). The data is color coded according to the three datasets: this work (blue circles), \citetalias{Adibekyan2024} (green squares) and \citetalias{Brinkman2024} (orange diamonds). The GALAH dataset is shown with black contours at 1 and 2 $\sigma$ confidence values.}
        \label{fig:GALAH}
    \end{figure}

    \paragraph*{Stellar parameters:}\ 
    our stellar characterization model, that ensures self-consistency across stellar parameters (e.g. {$T_{eff}, log(g), [Fe/H], M_{st}, R_{st}$}) with thorough inspection of each absorption line feature ([Fe/H], [Mg/H], [Si/H], [Ca/H], [Al/H], [Ni/H], [O/H]), yields Mg and Si abundances that are on average higher compared to \citetalias{Adibekyan2024} and \citetalias{Brinkman2024} (see Fig. \ref{fig:param_comp}).
    These larger Mg and Si abundances translate to lower Fe-MF$_{st}$ values. 
    {These differences primarily stem from our self-consistent methodology and stellar abundance retrieval pipeline, which result in different elemental abundances ([Fe/H], [Mg/H], [Si/H]) as well as stellar metallicity estimates.} 
    Moreover, we account for several sources of uncertainty, including a non-LTE treatment leading to larger errors in Fe-MF$_{st}$, which we deem more reliable. 
    This means that for most stars, our results are consistent within 1 $\sigma$ with \citetalias{Adibekyan2021}/\citetalias{Adibekyan2024} and \citetalias{Brinkman2024} stellar abundances, but not the other way around as the other two datasets reported lower, but perhaps underestimated, errors on abundance values ([Fe/H], [Mg/H], [Si/H]). 
    We illustrate these relationships between Fe-MF$_{st}$ and stellar abundances in Figure \ref{fig:GALAH}, with a comparison to the GALAH dataset.
    We observe a clear correlation between Fe-MF$_{st}$ and [Fe/H] for lower metallicity stars in the GALAH dataset.
    Thus, it is vital to understand whether the composition of rocky exoplanets around these stars is similar to or enhanced in comparison to the host star composition.
    However, inside the limited metallicity range of our host stars (-0.2<[Fe/H]<0.5), we find that none of the abundances are strongly correlated with the equivalent Fe-MF$_{st}$ composition as well as Mg/Si refractory ratio, opposite to \citetalias{Adibekyan2024} and \citetalias{Brinkman2024} data that have a negative slope between Fe-MF$_{st}$ vs Mg/Si.
    Furthermore, we find that our host stars have a wider range of Mg/Si ratios in comparison to others.
    This should not be overlooked in the case of stars, as changing Mg/Si ratio has a moderate effect on the total Fe-MF$_{st}$ budget, which is unlike the case for exoplanets where the Fe-MF$_{pl}$ is set by the M-R data and their large uncertainties \citep{Plotnykov2024}.
    
    \paragraph*{Exoplanet parameters:}\ 
    changes in exoplanet data are mainly driven by the updates in stellar radius and transit depth, that feed into the planetary radius inference. 
    For transit depth, we use the most precise measurements and make sure that each transit comes from the same instruments, similar to \citetalias{Brinkman2024}.
    Meanwhile, \citetalias{Adibekyan2021} and \citetalias{Adibekyan2024} do not use isochrone analysis to update stellar M-R and instead adopt the most recently reported exoplanet M-R measurements taken by the same group.
    The different planetary radii used by \citetalias{Adibekyan2021} ($\Delta R/R=4\%$) translate to an average $\Delta$Fe-MF$=0.1$ compared to our sample, but this difference is lower in the revised \citetalias{Adibekyan2024} dataset ($\Delta R/R=2\%$ and $\Delta$Fe-MF$=0.05$). 
    Moreover and similar to \citetalias{Adibekyan2024} and \citetalias{Brinkman2024}, with revised transit radii, we find a reduction of the number ultra dense exoplanets, where 3 out of 4 exoplanets that had previously been identified as super-Mercuries with Fe-MF$>0.6$ \citep{Plotnykov2020, Adibekyan2021} are no longer as iron-enriched, with K2-106~b changing the most (see section \ref{disc:SM}).
    In general though, our exoplanet and stellar M-R data is mostly consistent with other datasets within $1 \sigma$, except for 55-Cnc~e and K2-106~b (see Fig. \ref{fig:param_comp}, \ref{fig:mr_all}).
    
    \paragraph*{Outlier treatment:}\
    a selection of exoplanets that are on the verge of volatile-to-rocky boundary have a profound impact on setting the composition relationship and to solve this problem we have weighted each exoplanet according ($w_i=P_\mathrm{rocky}$, see Sec. \ref{correlation}).   
    As mentioned other studies instead employ non-statistical methods to address this issue: \citetalias{Adibekyan2021} considers the RTR planets to be rocky artificially increasing their precision, while \citetalias{Brinkman2024} allows for negative CMFs. 
    \citetalias{Adibekyan2024} instead uses a more comprehensive, but time-consuming way to model these straddling planets based on an interior structure model that allows for $\mathrm{H_2O}$ throughout the planet \citep{Luo2024}. 
    However, without atmospheric retrievals there is no guarantee that a water envelope is present or how much water has survived in the rocky interior (mantle or core), given that these exoplanets have high incident fluxes \citep{Fulton2017}.
    By assuming there is water in the interior of compact planets, the iron component increases to offset the added mass. Thus, assuming a rocky composition yields Fe-MF values that are on average 20 points lower, and represent a lower bound for iron content. For consistency, we recalculate the Fe-MF for the different datasets (\citetalias{Adibekyan2021}, \citetalias{Brinkman2024}, and \citetalias{Adibekyan2024}) to properly compare methodologies. 
    
\begin{figure}
        \centering
        \includegraphics[width=\linewidth]{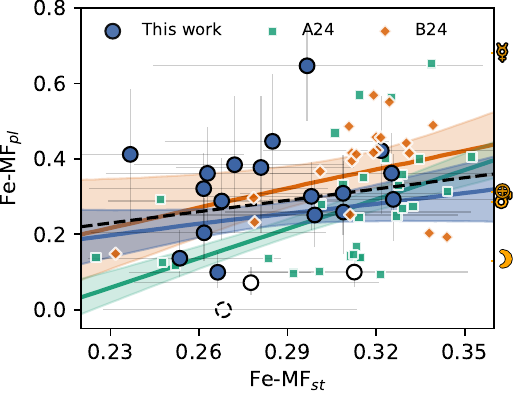}
        \caption{Comparison of stellar iron-mass fraction (Fe-MF$_{st}$ ) and planetary iron-mass fraction (Fe-MF$_{pl}$) using our methodology on different datasets: this work (blue circles), \citetalias{Adibekyan2024} (green squares) and \citetalias{Brinkman2024} (orange diamonds). The contours indicate best-fit $1 \sigma$ confidence interval. Note that we re-calculate both \citetalias{Adibekyan2024} and \citetalias{Brinkman2024} data using our interior and chemical model as well as our statistical approach yielding different relations to those proposed in the original studies.}
        \label{fig:femf_comparison}
    \end{figure}

    \paragraph*{Best-fit methods:}\
    in this study we use a Bayesian approach to fit a best-line between exoplanet and stellar data, instead of ODR used by both \citetalias{Adibekyan2021} and \citetalias{Brinkman2024}, that allows for asymmetrical errors in the data and better accounting of outliers, which are highly prevalent in these datasets.
    For the Fe-MF comparison, \citetalias{Adibekyan2024} applied different statistical methods including a Bayesian model ($m\sim3$), but it did not account for asymmetrical errors nor is the dataset weighted to account for outliers. 
    \citetalias{Brinkman2024} also applied OLS approach and obtained close to 1:1 relationship ($m=1.3$), which in conjunction with their steep relation from ODR they interpreted as there not being enough evidence of a robust correlation. 
    With a Bayesian approach, instead, we can inspect the posterior and identify any underlying discrepancies in the fit, thus quantifying the confidence in the relation as well as identify outliers, especially in light of a small data set. 
    Therefore, we apply our nominal weighted fitting model (see Sec. \ref{correlation}) to the \citetalias{Adibekyan2024} and \citetalias{Brinkman2024} datasets for its added statistical advantage.
    We find that \citetalias{Brinkman2024} dataset has a weaker slope to the one they report, which is closer to 1:1 $m=1.71^{+1.27}_{-1.46}$, while the \citetalias{Adibekyan2024} dataset has a slope that is slightly shallower $m=2.6\pm 0.6$ than their reported slope (see Fig. \ref{fig:femf_comparison}).
    This result indicates that both our and \citetalias{Brinkman2024} datasets are fundamentally different compared to that of \citetalias{Adibekyan2024}, which has a steeper slope.
    Furthermore, this is not attributed to the larger sample size of \citetalias{Adibekyan2024} (N=32) given the fact that when we limit our re-analysis of their data to the same exoplanets that are in our sample, we find the slope is still steep ($m=3.2 \pm 0.8$).
    Thus, it is most likely due to the self-consistent stellar and planetary analysis done in this work.

\subsection{Data needed to unequivocally test formation scenarios}
    We have seen so far that the systems with rocky exoplanets suggest a 1:1 relationship in the composition space between stars and their planets.
    {Meanwhile, we observe that the refractory ratios of exoplanets span a wider range compared to the host stars. This results in a best-fit slope between Fe-MF values to be unresolved, such that $m=0$ (no correlation) is within a one sigma confidence level, owing to the small sample size and large uncertainties in the data.
    In response to this ambiguity, we simulate a mock sample that in truth follows a perfect 1:1 Fe-MF relationship to (1) ascertain how likely the observed sample can be produced by random chance due to observation scatter and (2) to quantify the size of the sample needed to accurately recover a 1:1 slope, should it exist, assuming the current observational uncertainties in exoplanets. 
    }

\begin{figure}
        \centering
        \includegraphics[width=\linewidth]{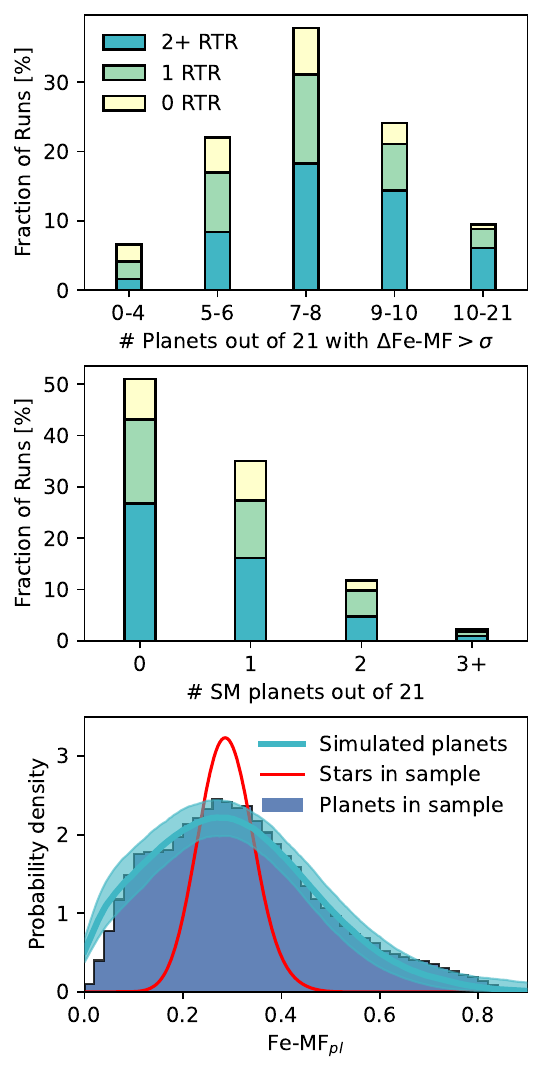}
        \caption{{Population statistics of simulated exoplanets (N$_{pl}=21$) given primordial origin and current measurement scatter, color coded according to the number of RTR-like ($\rho/\rho_\oplus<0.8$) planets present in each run. Top panel shows the number density of observing exoplanets with $\Delta \mathrm{Fe-MF} > \sigma$, middle panel highlights the likelihood of observing super-Mercuries (SM) and bottom panel corresponds to marginalized probability density of Fe-MF$_{pl}$ for all 21 planets in our and simulated sample as well as stars. The outcome of randomly observing 1 Mercury-like and 2-3 RTR-like exoplanets, which is similar to our dataset, occurs $\sim 15\%$ of the time.}}
        \label{fig:mock_data}
    \end{figure}

    {We use the GALAH dataset \citep{GALAH} to generate a population of stars and assume that each star hosts an exoplanet with some mass $\mathrm{M_{pl}} \sim\mathcal{U}(3,12) \ \mathrm{M_\oplus}$ and corresponding radius such that the refractory ratios are exactly the same as the host star. Subsequently, we apply the current observational error scatter to this mock population ($\text{Fe-MF}_{\text{error}} \sim 0.03 \text{ and } 0.12$, for stars and exoplanets), and fit this Fe-MF star-planet mock data to a line to recover the slope and intercept.
    We repeat this process 1000 times to find the variance in these parameters.
    We find that the generated mock populations are fairly similar to our self-consistent exoplanet population.  Our results for simulations with 21 planets in the sample (N$_{pl}=21$) are shown in Figure  \ref{fig:mock_data}.
    The most likely outcome is a population with no super-Mercuries -- comprising half of the outcomes -- although 1 or even 2 are still possible, accounting for 35$\%$ and 12$\%$ of the outcomes, respectively. 
    We find instead that having $\leq1$ or $\geq 2$ RTR-like planets is equally probable and irregardless of the number of outliers, we always recover a slope that is consistent with 1 and 0.
    Thus, the likelihood of randomly observing a population with 1 super-Mercury and 2-3 RTR-like exoplanets out of 21, which is what we have in our sample, is $\sim 15\%$.  
    The results from this mock population indicate that the larger range observed in the composition of exoplanets compared to that of stars may be explained, at the $\sim 15\%$ level, by the scatter produced due to observation uncertainty. It also means, however, it is entirely possible that the spread in planetary composition is real and reflects formation pathways away from a primordial origin. 
    A larger sample and especially the prevalence or absence of super-Mercuries is the path to determine whether the spread in composition is real or an artifact of errors.
    }
    
\begin{figure}
        \centering
        \includegraphics[width=\linewidth]{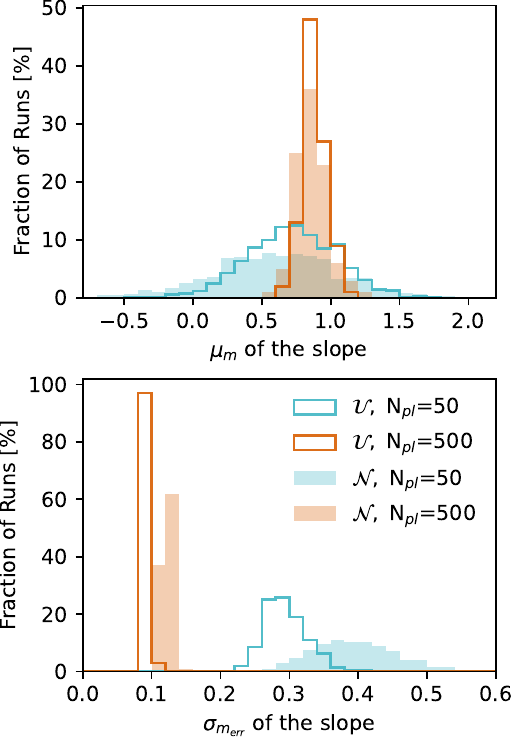}
        \caption{Distributions of the recovered slope (m) means (top panel) and $\sigma$ error (bottom panel) from each best-fit result, given a sample with 50 (1000 runs) or 500 planets (100 runs). Provided that the initial set (before applying observational error) follows 1:1 relationship planets and their host that we sample from GALAH population ($\mathcal{N}$, solid histograms) or from uniformly spaced stellar sample ($\mathcal{U}$, transparent histograms), Fe-MF$_{st} \sim \mathcal{U}(0.2,0.4)$.}
        \label{fig:mock_m}
    \end{figure}

    Next we examine the sample size needed to determine unequivocally a 1:1 relation given the current level of mass and radius certainty. Unsurprisingly but unfortunately, we find that a sample size of N$_{pl}\leq50$, like the one we have, is unlikely to recover the true relationship ($m=1$) given the large uncertainty in the data --  $\sim40\%$ have m=0 within $2\sigma$ confidence (see Fig. \ref{fig:mock_m}).
    However, we find it is possible to robustly recover the correct best-fit line back, at 95$\%$ confidence when there are 150 planets.
    For a higher confidence level (3 $\sigma$), we find that for N=150 there are around a quarter of runs with $m<0$ in the posterior.
    This means that to confidently constrain the composition using only Fe-MF of planets and stars, a sample that of at least 7 times the current size is needed.
    The good news is that PLATO is expected to deliver about 1000 planets with radii of $0.8-1.25 R_\oplus$ after the first four mission years \citep{PLATO}.

    In the meantime, one way to improve the outcome is to focus on stars that are on the fringes of the Fe-MF space. 
    For example, if we sample the stellar composition uniformly, such that there is equal distribution of Fe-MF$_{st}$ from low to high, fewer planets are needed to constrain the relationship. 
    We reach 95$\%$ confidence with N$\sim 60$.
    Thus, an efficient approach to mining the exoplanet rocky population is to target a diverse stellar group, which increases the lever arm in the correlation. 
    Furthermore, meaningful correlations will only be possible with an adequate treatment of outliers and the consideration of subpopulations.  
    
\section{Discussion}
\subsection{Super-Mercuries}
\label{disc:SM}
    We have shown the need to carefully select planetary masses and radius and in particular guarantee consistency in stellar parameters to ensure consistency in planetary M-R data. 
    One of the shortcomings of using inconsistent data was the apparent existence of a group of super-Mercuries \citet{Plotnykov2020, Adibekyan2021, Schulze2021}) with implications on the existence of our own Mercury. 
    In our solar system, Mercury holds a unique position as the innermost planet with an unusually large amount of iron, with Fe-MF$_{pl} \sim 0.67$ \citet{Hauck2013} or 2-times enriched with respect to solar, making it an outlier compared to Earth, Venus and Mars which have similar refractory ratios to that of the Sun \citep{Anders1989,Grevesse2007,Lodders2009,Asplund2009}. There have been different proposed formation mechanisms \citep{Ebel2011,Hubbard2014,Cuello2016,Kruss2020,Aguichine2020,Johansen2022} to explain Mercury's composition with giant impacts being the leading theory \citep{Benz2007} albeit not without problems \citep{Chau2018,Clement2021,Franco2022,Scora2024}. 
    It remains to be seen how unique its formation is. 
    Thus, super-Mercuries are prime targets for understanding planetary formation and evolution theories. 

    In our updated dataset, the exoplanet M-R are revised such that we observe fewer super-Mercuries--only Kepler-107~c remains--and there is no evidence for a sub-population as seen by \citetalias{Adibekyan2021}, in agreement with \citetalias{Brinkman2024}.
    This suggest the prevalence of fewer super-Mercuries in general, which we attribute to our homogeneous modeling and updates to the transit and RV observations.
    Nevertheless, we still find a few exoplanets--K2-38~b, K2-229~b and K2-291~b--that have relative iron-enrichment compared to their host star ($\Delta$Fe-MF$\sim 0.2$).
    Some of these were previously thought super-Mercuries and would be important to follow up, given that there is no current formation theory to explain even this smaller iron-enrichment for massive rocky planets \citep{Scora2020,Scora2024}. 

\subsection{Multi-Planet systems}
    Host stars that have multiple rocky planets in orbit are also prime targets for evaluating how the composition gest sent during formation and early evolution. 
    Evaluating the discrepancies between the interior composition of 2 or more planets around the same star is a fruitful way to uncover formation pathways.   

    \begin{figure}
        \centering
        \includegraphics[width=\linewidth]{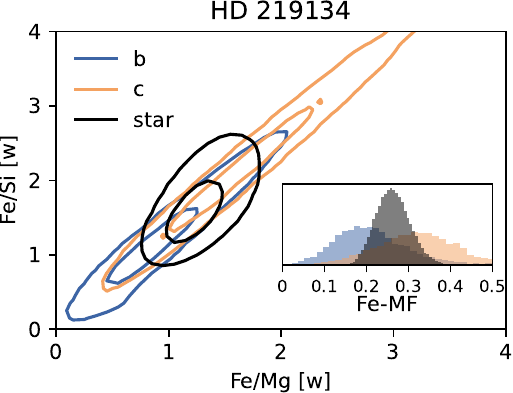}
        \caption{Comparison of the host star HD~219134 (black) refractory ratios to that of planets b (blue) and c (orange). The inlet plot shows the comparison in iron-mass fraction space.}
        \label{fig:multi}
    \end{figure}
    
    In our sample, HD~219134 is the only system with two exoplanets. 
    We investigate the similarities between planet b and c composition and their host star as an example.
    We find that both exoplanets have refractory ratios and Fe-MFs that agree within 1 $\sigma$ with the host star, but differ by more than 1 $\sigma$ confidence level between each other (see Fig. \ref{fig:multi}).
    It appears that planet b has less, while planet c has more iron compared to the host star, consistent with a formation process that enhances the iron content in planets closer to their star. 
    However, no definite conclusions can be drawn given the large errors that cause the posterior compositions to overlap widely. 
    With better data and more systems, this line of inquiry can offer more definite answers.

\subsection{Role of galactic evolution}
\label{disc:Age_Fe}
    Galactic Chemical Evolution (GCE) governs the metal content we observe in the stellar photospheres. 
    In particular, the abundance ratios of $\alpha$ elements (Mg, Si, Ca, Ti) to iron ([$\alpha$/Fe]) is simply put, a direct measurement of the different contributions of core-collapse supernovae (which favors the production of $\alpha$ elements \citealt{Timmes1995,Kobayashi2006}) versus the contribution of thermonuclear explosions (type Ia supernovae - which contribute more significantly to the production of iron peak elements - Cr, Mn, Fe, Co, Ni, Cu, Zn \citealt{Kobayashi2009,Kobayashi2020}).
    The observed decrease in [$\alpha$/Fe] as stellar metallicity ([Fe/H]) increases is a direct result of the time-delay difference between core collapse and type Ia supernovae. 
    Therefore, in a nutshell, most $\alpha$ elements are produced by massive stars (M$_{st}>8$M$_\odot$) within 20-30 Myrs, while Fe is produced primarily on a much longer timescale \citep[up to $\sim$ 1 Gyr, see][for a GCE review]{Kobayashi:2020ApJ...900..179K}.
    From the perspective of GCE, [$\alpha$/Fe] can, therefore, be a proxy of how late or early a star has formed. 
    It is important to note, that this is a simplified view that does not account for the complex mixing processes of the interstellar medium (ISM) or the time delay required for the ISM to homogenize, prior to formation of new generation of stars, nor the specific details of the chemical evolution of the galactic substructure where the star was formed (i.e. thin disk, thick disk, bulge, halo).
    
    In these simple terms though and referring to trends applicable for a large population of stars, most stars formed in the thick disk are $\alpha$-rich and possibly older, compared to thin disk stars, which mostly have Sun-like [$\alpha$/Fe] ratios \citep{Gilmore1983, Matteucci:1989epg..conf..297M,Fuhrmann1998, Kobayashi:2020ApJ...900..179K}.
    The expectation then is that the Fe-MF$_{pl}$ of exoplanets and stellar Fe-MF$_{st}$ values, which is directly proportional to [$\alpha$/Fe] ratios, could have a correlation with stellar age {or be affected by GCE as shown and argued by \citet{Weeks2025,Behmard2025}}.

    \begin{figure}
        \centering
        \includegraphics[width=\linewidth]{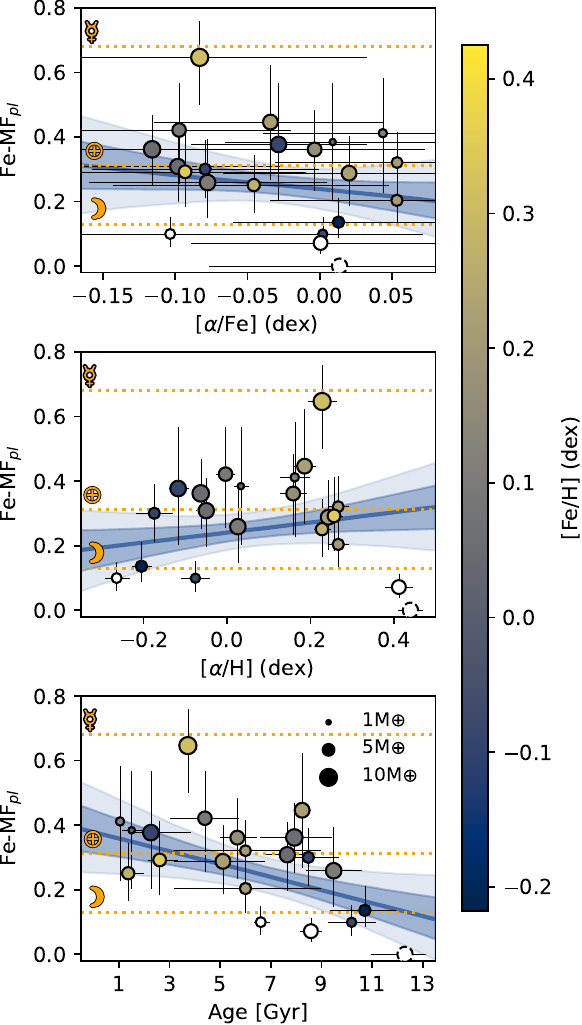}
        \caption{Planet's iron-mass fraction (Fe-MF$_{pl}$) as a function of alpha elements to iron ratio ([$\alpha$/Fe], top panel), alpha elements abundance ([$\alpha$/H], middle panel) and  stellar age calculated with isochrone analysis (bottom panel). Each point is color coded by the stellar [Fe/H] abundances and the marker size represents the observed planet mass.  We use the same color scheme for planets that are above purely rocky threshold as before and show in blue our best-fit result including the 1 and 2 $\sigma$ values.}
        \label{fig:Age}
    \end{figure}

    We investigate this and find that in our sample neither the Fe-MF$_{st}$ nor individual abundances of stars are correlated with age.  In contrast for planets, however, we find a negative slope between Fe-MF$_{pl}$ and age of $m=-0.02 \pm 0.01$ (see Figure \ref{fig:Age}), in line with iron enrichment in the ISM over time {and in agreement with \citet{Weeks2025} results}.
    {The former outcome is unsurprising, given that the stars in our sample span a limited range in both [Fe/H] and [$\alpha$/Fe] values, consistent with their common Galactic substructure origin. 
    Consequently, the absence of a correlation likely reflects limited precision as the intrinsic stellar differences are expected to be small \citep{Lorenzo2018}.
    Furthermore, the stellar ages in our sample are derived from isochrone analysis alone, which does not account for systematics and thus our stellar ages require benchmarking to other methodologies for robustness. 
    For example, using chemical clock analysis \citep{Shejeelammal2024}, we find that 55 Cnc has a drastically different stellar age of $3.8\pm 2$ Gyr in comparison to our isochrone result of 12.30$^{+0.82}_{-1.34}$ Gyr.
    These factors introduce intrinsic challenges of measuring stellar ages, we refer the reader to \citet{Reggiani2022, Reggiani2024} for a detailed description of our methodology and \citet{soderblom2010} for a comprehensive review of the difficulties in estimating the ages of field stars.}
  
    {For the latter outcome on planets, there is possibly also a source of strong bias. Radial-velocity jitter and hence the observed $M_{pl}$ uncertainty, scales with stellar age \citep{Brems2019}, with younger stars exhibiting higher jitter. Meaning, there may exists a strong bias toward planets with lower densities and lower Fe-MF, since for a given radius ($R_{pl}=1.6R_\oplus$) planets that straddle the RTR boundary (CMF$\sim0$) are only detectable around older ($\gtrsim$ 5 Gyr) host stars (see \citetalias{paperI} for details).
    Thus, these linear fits obtained with a few dozen planetary systems are most likely an overestimate of the correlation strength.
    Instead of stellar ages, hence, we compare the host star [$\alpha/Fe$] as well as [$\alpha/H$] abundances to the Fe-MF$_{pl}$, which we discover to not have as robust of a linear relationship.
    We find that Fe-MF$_{pl}$ exhibits a negative correlation with [$\alpha$/Fe] ($m = -0.4 \pm 0.5$), while its relationship with [$\alpha$/H] yields a positive slope ($m = 0.15 \pm 0.14$).
    In both cases, however, no correlation ($m = 0$) is consistent within $\sim 1\sigma$.}
    These findings raise interesting questions about whether planetary composition changes with galactic evolution and if older stars are more likely to host certain types of planets. 
    Necessary improvements for this test require accurate and robust stellar ages or kinematics to determine thin/thick disk provenance (outside the scope of our paper) as well as a sample size that includes stars at a wider range of abundances ([Fe/H]$<-0.2$ or [Fe/H]$>0.4$, which will depend on upcoming planetary discovery missions).

\section{Conclusions}
    In this study, we employ a state-of-the-art interior structure model alongside a rigorous statistical approach to reanalyze exoplanet and stellar data, that is ensured to be self-consistent, physically sound and homogeneous. Our analysis encompasses 21 systems and 22 exoplanets, aiming to uncover potential relationships. We carefully included planets straddling the volatile-rocky boundary by associating a weight proportional to the area in the M-R space that overlaps the rocky region. This is a summary of our key findings:
\begin{itemize}
    \item We uncover a near 1:1 relationship between planets and stars in Fe-MF space, Fe-MF$_{pl} = 0.94^{+1.02}_{-1.07} \cdot$ Fe-MF$_{st} - 0.02^{+0.31}_{-0.29}$. This result suggests a primordial origin composition link when looking at the ensemble of planets. We find this relationship is consistent with data from \citetalias{Brinkman2024} when the planets that straddle the volatile-rocky composition boundary are properly treated, while it is in disagreement with data from \citetalias{Adibekyan2021}/\citetalias{Adibekyan2024}.
    \item     The new data reduces the number of super-Mercuries from 4 in our sample to only Kepler-107~c, although some of these previously labeled super-Mercuries (K2-38~b, K2-229~b, K2-291~b) remain iron enriched beyond what planet formation models can predict. 
    \item Despite a linear relationship, planetary data spans a larger compositional space compared to stars. There are 5 exoplanets with an Fe-MF different than their host star beyond $1 \sigma$ confidence level. Compared to our earlier work, the self-consistent treatment of the stellar parameters and updates to transit radius yield more exoplanets that are closer to an Earth-like or star-like composition. 
    \item  {However, we also find with mock data analysis that the observed spread in exoplanet composition is within expectations when considering observation uncertainties that have a primordial origin, including a $15\%$ chance of having one super-Mercury and two to three underdense exoplanets out of 21. In other words, the scatter in planet composition can be an artifact of observational scatter around a stellar composition.} 
    \item     To determine the size of the sample needed to truly confirm the reliability of a primordial origin suggested by the 1:1 relation, we use mock data. We find that a sample of at least 150 planets is needed. This can be improved to only 60 planets by targeting metal poor and metal rich stars and thus increasing the lever arm in the correlation.
    \item     An avenue for further improvement consists of obtaining more data on multi-planet systems as they share the same star. 
    \item {Lastly, we find that the relationship between Fe-MF$_{pl}$ and both stellar [$\alpha$/Fe] and [$\alpha$/H]  are strongly consistent with a null correlation result (m=0). This stands in contrast to the negative relationship between Fe-MF$_{pl}$ and the stellar age. Thus, we caution against deriving conclusions about compositional trends with age without a thorough analysis of systematic errors.}

\end{itemize}

\begin{acknowledgments}
    This work has been partially funded by the Natural Sciences and Engineering Research Council of Canada (Grant RGPIN-2021-02706). We would like to acknowledge that our work was performed on land traditionally inhabited by the Wendat, the Anishnaabeg, Haudenosaunee, Metis and the Mississaugas of the New Credit First Nation.
\end{acknowledgments}

\bibliography{bibliography}{}
\bibliographystyle{aasjournal}

\begin{figure*}[h!]
    \centering
    \includegraphics[width=\linewidth]{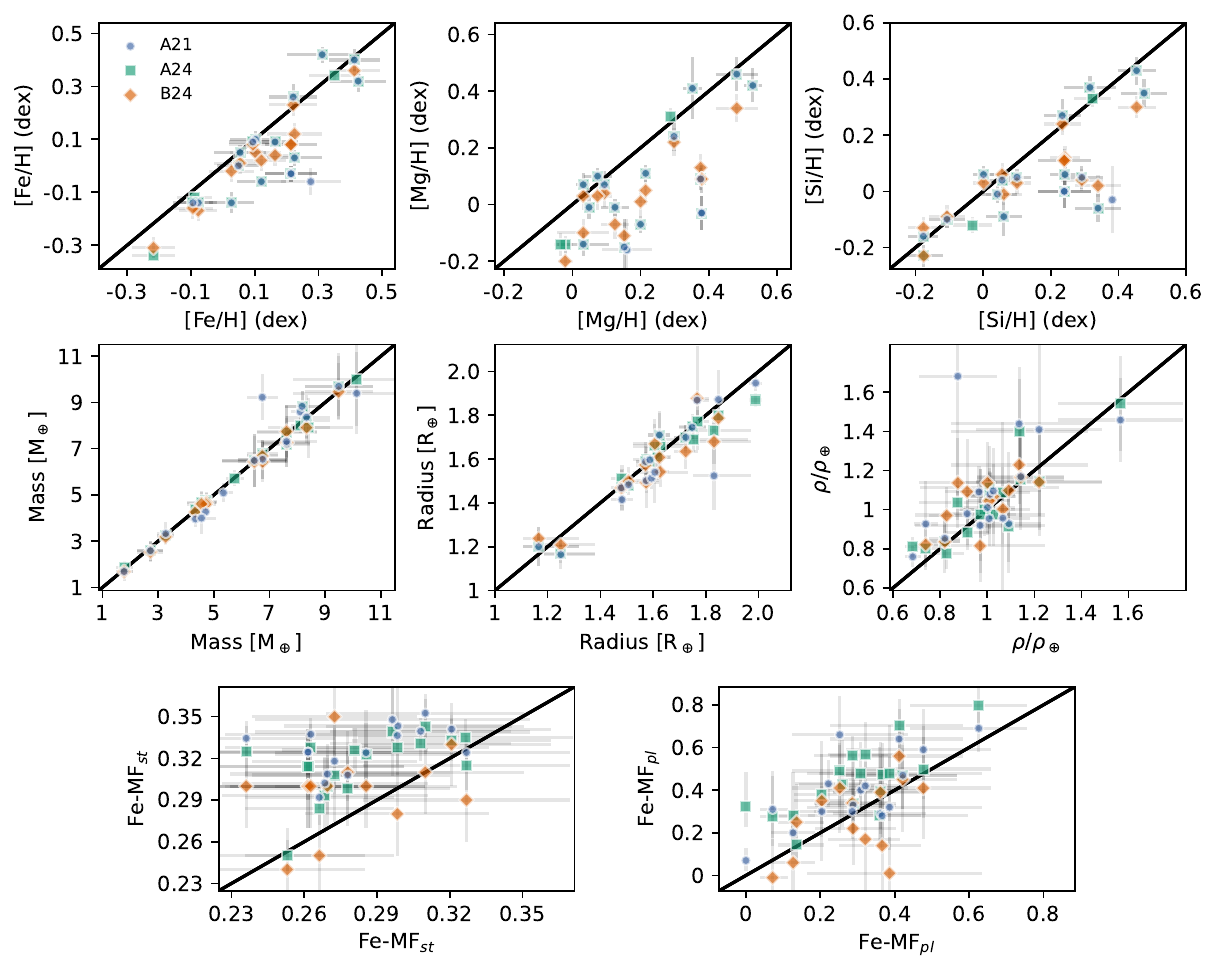}
    \caption{Comparison of this work stellar and planetary data to \citetalias{Adibekyan2021}, \citetalias{Adibekyan2024} and \citetalias{Brinkman2024} data. Our data is on the x-axis, while other works are on the y-axis. We show stellar abundances obtained with non-LTE corrections and corresponding Fe-MF$_{st}$ values. The Fe-MF$_{st}$ and Fe-MF$_{pl}$ shown here are those reported by the respective works, which assume different interior and chemical models.} 
    \label{fig:param_comp}
\end{figure*}

\begin{figure*}
    \centering
    \includegraphics[width=\linewidth]{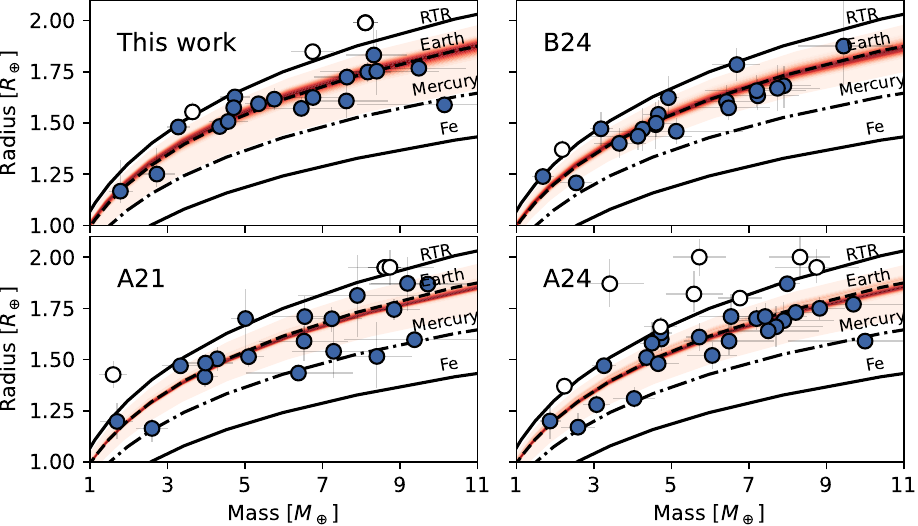}
    \caption{Comparison of planetary masses and radii, for this work, \citetalias{Adibekyan2021}, \citetalias{Adibekyan2024} and \citetalias{Brinkman2024} datasets. The red contours show the population of host star planet equivalent values for each dataset. Exoplanets above the RTR, which require volatiles, are indicated as white circles.}
    \label{fig:mr_all}
\end{figure*}

\begin{figure*}
    \centering
    \includegraphics[width=\linewidth,height=9in]{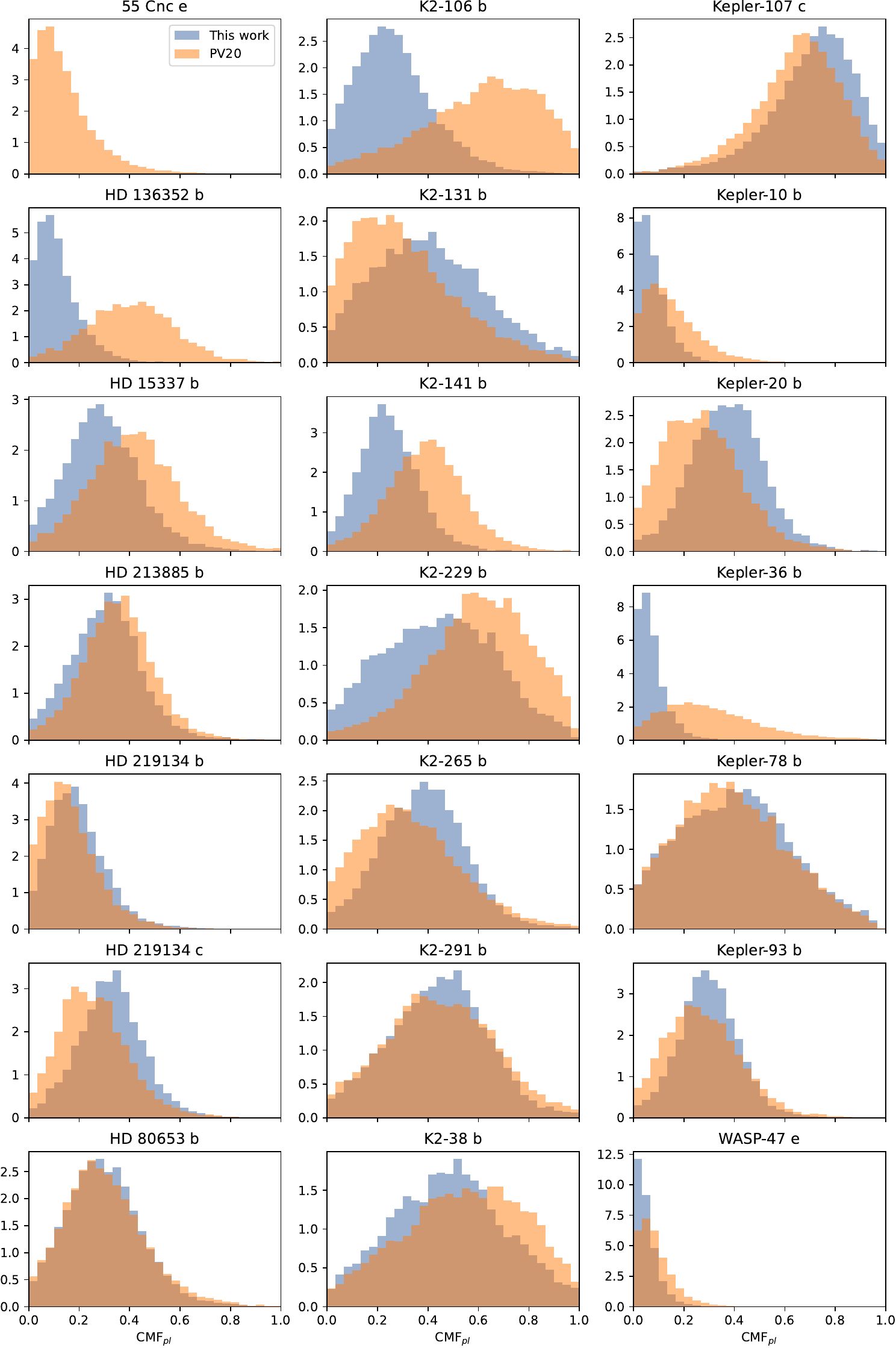}
    \caption{Posterior probability density of CMF obtained from our MCMC simulations using exoplanet homogeneous M-R data from this work, compared to the heterogeneous sample used in \citet{Plotnykov2020}. The new data has a general trend towards lower CMF values compared to the older values, with K2-106~b's CMF changing the most.}
    \label{fig:data_full}
\end{figure*}

\appendix
    This section presents the comprehensive planetary-star composition results of our study (see Tab. \ref{tab:planet},\ref{tab:star}, \ref{tab:coef_ext}), alongside a detailed graphical comparison with \citetalias{Adibekyan2021}, \citetalias{Adibekyan2024} and \citetalias{Brinkman2024} data (see Fig. \ref{fig:param_comp},\ref{fig:mr_all}).
    Furthermore, we show inferred planetary CMF posterior probability density in comparison to the \citet{Plotnykov2020} data, Fig. \ref{fig:data_full}.

\begin{table*}[]
    \centering
    \renewcommand{\arraystretch}{1.5} 
    \begin{tabular}{cccccccccc}
        Name  & M$^\dagger$ (M$_\oplus$) & R$^\dagger$ (R$_\oplus$) & $\frac{\rho}{\rho_\oplus}$& CMF$^\ddagger$ & CMF & Fe-MF & $\frac{Fe}{Mg}$ (w) & $\frac{Fe}{Si}$ (w) & $\frac{Fe+Ni}{Si+Mg}$ (w) \\ \hline \hline
55~Cnc~e & - & - & 0.68$^{+0.03}_{-0.03}$ & 11.7$^{+13.0}_{-7.5}$ & - & - & - & - & - \\ \hline 
HD~136352~b & 4.96$^{+0.32}_{-0.28}$ & 1.61$^{+0.02}_{-0.03}$ & 0.83$^{+0.08}_{-0.07}$ & 40.6$^{+16.7}_{-17.2}$ & 10.0$^{+9.5}_{-6.1}$ & 13.6$^{+7.4}_{-5.0}$ & 0.58$^{+0.38}_{-0.24}$ & 0.78$^{+0.53}_{-0.36}$ & 0.35$^{+0.24}_{-0.15}$\\ \hline 
HD~15337~b & 7.73$^{+0.74}_{-0.73}$ & 1.72$^{+0.04}_{-0.04}$ & 1.00$^{+0.12}_{-0.11}$ & 41.2$^{+17.3}_{-16.9}$ & 28.8$^{+14.6}_{-13.7}$ & 28.8$^{+11.4}_{-10.2}$ & 1.52$^{+1.02}_{-0.67}$ & 1.97$^{+1.31}_{-0.87}$ & 0.93$^{+0.62}_{-0.41}$\\ \hline 
HD~213885~b & 8.18$^{+0.64}_{-0.59}$ & 1.74$^{+0.05}_{-0.05}$ & 1.01$^{+0.11}_{-0.1}$ & 35.4$^{+13.2}_{-13.5}$ & 30.8$^{+12.7}_{-14.2}$ & 30.9$^{+10.1}_{-11.1}$ & 1.88$^{+1.02}_{-0.84}$ & 1.81$^{+0.98}_{-0.81}$ & 0.98$^{+0.53}_{-0.44}$\\ \hline 
HD~219134~b & 4.72$^{+0.18}_{-0.17}$ & 1.56$^{+0.03}_{-0.04}$ & 0.91$^{+0.1}_{-0.08}$ & 15.0$^{+12.0}_{-9.0}$ & 18.3$^{+11.9}_{-9.3}$ & 20.4$^{+9.3}_{-7.4}$ & 0.95$^{+0.62}_{-0.4}$ & 1.24$^{+0.8}_{-0.54}$ & 0.58$^{+0.38}_{-0.25}$\\ \hline 
HD~219134~c & 4.35$^{+0.19}_{-0.2}$ & 1.48$^{+0.04}_{-0.04}$ & 1.02$^{+0.1}_{-0.09}$ & 24.9$^{+14.1}_{-11.9}$ & 32.6$^{+12.3}_{-12.7}$ & 32.1$^{+9.4}_{-10.0}$ & 1.76$^{+0.89}_{-0.71}$ & 2.29$^{+1.16}_{-0.93}$ & 1.07$^{+0.54}_{-0.43}$\\ \hline 
HD~80653~b & 5.78$^{+0.35}_{-0.32}$ & 1.60$^{+0.05}_{-0.05}$ & 0.98$^{+0.12}_{-0.1}$ & 28.8$^{+16.9}_{-14.0}$ & 29.1$^{+15.2}_{-14.1}$ & 29.2$^{+12.0}_{-11.0}$ & 1.82$^{+1.31}_{-0.86}$ & 1.59$^{+1.14}_{-0.75}$ & 0.91$^{+0.65}_{-0.43}$\\ \hline 
K2-106~b & 8.51$^{+0.72}_{-0.68}$ & 1.78$^{+0.06}_{-0.06}$ & 0.88$^{+0.16}_{-0.17}$ & 63.5$^{+20.5}_{-26.7}$ & 25.0$^{+16.7}_{-13.3}$ & 25.9$^{+13.4}_{-11.1}$ & 1.46$^{+1.29}_{-0.73}$ & 1.37$^{+1.21}_{-0.69}$ & 0.75$^{+0.67}_{-0.38}$\\ \hline 
K2-131~b & 8.54$^{+1.57}_{-1.37}$ & 1.71$^{+0.1}_{-0.1}$ & 0.99$^{+0.31}_{-0.28}$ & 27.8$^{+24.2}_{-16.8}$ & 39.3$^{+23.7}_{-21.5}$ & 37.7$^{+19.0}_{-17.3}$ & 2.57$^{+3.18}_{-1.51}$ & 2.41$^{+2.98}_{-1.41}$ & 1.32$^{+1.64}_{-0.78}$\\ \hline 
K2-141~b & 5.36$^{+0.33}_{-0.31}$ & 1.59$^{+0.03}_{-0.03}$ & 0.96$^{+0.08}_{-0.08}$ & 39.3$^{+14.2}_{-15.0}$ & 23.7$^{+12.0}_{-10.4}$ & 25.1$^{+9.3}_{-8.5}$ & 1.81$^{+0.98}_{-0.73}$ & 1.06$^{+0.67}_{-0.45}$ & 0.73$^{+0.45}_{-0.31}$\\ \hline 
K2-229~b & 2.74$^{+0.41}_{-0.4}$ & 1.26$^{+0.07}_{-0.07}$ & 1.13$^{+0.3}_{-0.29}$ & 60.8$^{+19.6}_{-21.8}$ & 44.9$^{+22.3}_{-24.2}$ & 41.2$^{+17.2}_{-18.5}$ & 3.26$^{+4.16}_{-2.03}$ & 2.80$^{+2.95}_{-1.76}$ & 1.63$^{+1.83}_{-0.97}$\\ \hline 
K2-265~b & 6.70$^{+0.85}_{-0.8}$ & 1.62$^{+0.04}_{-0.04}$ & 1.09$^{+0.14}_{-0.13}$ & 31.4$^{+21.8}_{-17.7}$ & 38.2$^{+15.9}_{-16.5}$ & 36.1$^{+12.3}_{-12.8}$ & 2.64$^{+1.86}_{-1.31}$ & 2.11$^{+1.49}_{-1.03}$ & 1.28$^{+0.9}_{-0.63}$\\ \hline 
K2-291~b & 6.38$^{+1.08}_{-0.98}$ & 1.57$^{+0.04}_{-0.04}$ & 1.14$^{+0.19}_{-0.18}$ & 44.2$^{+22.1}_{-21.3}$ & 44.6$^{+18.2}_{-20.6}$ & 42.2$^{+14.5}_{-16.8}$ & 2.97$^{+2.47}_{-1.59}$ & 2.94$^{+2.44}_{-1.57}$ & 1.57$^{+1.3}_{-0.84}$\\ \hline 
K2-38~b & 7.50$^{+1.16}_{-1.01}$ & 1.62$^{+0.09}_{-0.08}$ & 1.21$^{+0.28}_{-0.26}$ & 55.5$^{+23.8}_{-26.6}$ & 48.6$^{+22.9}_{-23.1}$ & 44.6$^{+17.8}_{-18.0}$ & 3.29$^{+3.69}_{-1.84}$ & 3.62$^{+3.7}_{-2.03}$ & 1.86$^{+2.07}_{-1.04}$\\ \hline 
Kepler-107~c & 9.09$^{+1.86}_{-1.8}$ & 1.59$^{+0.02}_{-0.02}$ & 1.57$^{+0.25}_{-0.27}$ & 66.2$^{+15.1}_{-18.5}$ & 72.7$^{+13.8}_{-18.5}$ & 64.7$^{+11.3}_{-14.6}$ & 7.84$^{+9.32}_{-3.69}$ & 7.63$^{+5.4}_{-3.35}$ & 4.19$^{+3.57}_{-1.93}$\\ \hline 
Kepler-10~b & 3.37$^{+0.2}_{-0.15}$ & 1.47$^{+0.02}_{-0.02}$ & 0.82$^{+0.07}_{-0.07}$ & 13.1$^{+12.7}_{-8.0}$ & 6.2$^{+6.6}_{-4.0}$ & 9.9$^{+5.4}_{-4.2}$ & 0.39$^{+0.24}_{-0.18}$ & 0.58$^{+0.43}_{-0.29}$ & 0.24$^{+0.17}_{-0.11}$\\ \hline 
Kepler-20~b & 9.41$^{+1.13}_{-1.02}$ & 1.77$^{+0.03}_{-0.02}$ & 1.07$^{+0.12}_{-0.12}$ & 26.3$^{+16.1}_{-14.3}$ & 37.2$^{+13.9}_{-14.2}$ & 36.2$^{+10.6}_{-11.0}$ & 2.34$^{+1.34}_{-0.94}$ & 2.38$^{+1.37}_{-0.96}$ & 1.26$^{+0.72}_{-0.51}$\\ \hline 
Kepler-36~b & 3.67$^{+0.09}_{-0.09}$ & 1.51$^{+0.02}_{-0.02}$ & 0.77$^{+0.06}_{-0.07}$ & 28.1$^{+21.9}_{-15.9}$ & 5.9$^{+6.1}_{-3.7}$ & 10.0$^{+5.1}_{-4.0}$ & 0.44$^{+0.29}_{-0.19}$ & 0.42$^{+0.3}_{-0.19}$ & 0.22$^{+0.15}_{-0.1}$\\ \hline 
Kepler-78~b & 1.82$^{+0.25}_{-0.26}$ & 1.14$^{+0.07}_{-0.06}$ & 0.97$^{+0.28}_{-0.29}$ & 37.9$^{+23.9}_{-20.3}$ & 40.4$^{+22.5}_{-22.6}$ & 38.4$^{+18.2}_{-18.2}$ & 2.37$^{+2.67}_{-1.41}$ & 2.77$^{+3.11}_{-1.65}$ & 1.37$^{+1.54}_{-0.81}$\\ \hline 
Kepler-93~b & 4.52$^{+0.48}_{-0.41}$ & 1.51$^{+0.01}_{-0.01}$ & 1.01$^{+0.08}_{-0.09}$ & 26.7$^{+15.8}_{-13.7}$ & 29.5$^{+11.7}_{-11.0}$ & 30.0$^{+9.0}_{-9.0}$ & 1.47$^{+0.71}_{-0.56}$ & 2.25$^{+1.1}_{-0.85}$ & 0.94$^{+0.46}_{-0.36}$\\ \hline 
WASP-47~e & 7.69$^{+0.35}_{-0.33}$ & 1.83$^{+0.02}_{-0.02}$ & 0.74$^{+0.05}_{-0.05}$ & 7.4$^{+7.5}_{-4.7}$ & 4.2$^{+4.9}_{-2.9}$ & 7.2$^{+4.1}_{-3.3}$ & 0.31$^{+0.2}_{-0.15}$ & 0.29$^{+0.21}_{-0.15}$ & 0.16$^{+0.11}_{-0.08}$\\ \hline 
    \end{tabular} 
    \\
    \begin{flushleft}
    $\dagger - $ Posterior results for Mass and Radius data, may be different from the observational data owing to sampling with MCMC as some distributions are outside the rocky region (RTR). \\
    $\ddagger - $ Planetary CMF data from \citet{Plotnykov2020}.
    \end{flushleft}
    \caption{Exoplanets parameters and composition results. The CMF and Fe-MF results are reported with respect to the percentage points (wt$\%$) and not as a fraction. }
    \label{tab:planet}
\end{table*}

\begin{table*}[]
    \centering
    \renewcommand{\arraystretch}{1.5} 
    \begin{tabular}{cccccccc}
        Name & M (M$_{\odot}$) & R (R$_{\odot}$) & [Fe/H] (dex) & [Mg/H] (dex) & [Si/H] (dex) & Fe-MF$_{st}$ & Age (Gyr) \\ \hline \hline
55~Cnc & 0.92$^{+0.01}_{-0.01}$ & 1.00$^{+0.01}_{-0.01}$ & 0.42$^{+0.09}_{-0.09}$ & 0.53$^{+0.03}_{-0.03}$ & 0.48$^{+0.07}_{-0.07}$ & 26.8$^{+4.5}_{-4.1}$ & 12.30$^{+0.82}_{-1.34}$\\ \hline 
HD~136352 & 0.88$^{+0.03}_{-0.02}$ & 1.03$^{+0.01}_{-0.01}$ & -0.22$^{+0.07}_{-0.07}$ & -0.02$^{+0.02}_{-0.02}$ & -0.18$^{+0.06}_{-0.06}$ & 25.3$^{+3.4}_{-3.1}$ & 10.71$^{+1.23}_{-1.37}$\\ \hline 
HD~15337 & 0.90$^{+0.02}_{-0.02}$ & 0.85$^{+0.01}_{-0.01}$ & 0.23$^{+0.08}_{-0.09}$ & 0.38$^{+0.02}_{-0.02}$ & 0.24$^{+0.05}_{-0.04}$ & 26.8$^{+4.2}_{-3.9}$ & 5.10$^{+1.66}_{-1.92}$\\ \hline 
HD~213885 & 0.96$^{+0.02}_{-0.02}$ & 1.10$^{+0.01}_{-0.01}$ & 0.05$^{+0.07}_{-0.07}$ & 0.05$^{+0.01}_{-0.01}$ & 0.04$^{+0.06}_{-0.06}$ & 30.9$^{+3.7}_{-3.5}$ & 7.65$^{+0.89}_{-0.83}$\\ \hline 
HD~219134 & 0.80$^{+0.02}_{-0.02}$ & 0.76$^{+0.01}_{-0.01}$ & 0.21$^{+0.08}_{-0.08}$ & 0.38$^{+0.01}_{-0.01}$ & 0.24$^{+0.08}_{-0.08}$ & 26.2$^{+4.1}_{-3.8}$ & 6.00$^{+2.97}_{-2.82}$\\ \hline 
HD~219134 & 0.80$^{+0.02}_{-0.02}$ & 0.76$^{+0.01}_{-0.01}$ & 0.21$^{+0.08}_{-0.08}$ & 0.38$^{+0.01}_{-0.01}$ & 0.24$^{+0.08}_{-0.08}$ & 26.2$^{+4.2}_{-3.9}$ & 6.00$^{+2.97}_{-2.82}$\\ \hline 
HD~80653 & 1.19$^{+0.01}_{-0.02}$ & 1.22$^{+0.01}_{-0.01}$ & 0.35$^{+0.08}_{-0.08}$ & 0.29$^{+0.01}_{-0.01}$ & 0.32$^{+0.05}_{-0.05}$ & 32.6$^{+4.7}_{-4.2}$ & 2.61$^{+0.71}_{-0.47}$\\ \hline 
K2-106 & 0.93$^{+0.01}_{-0.02}$ & 0.99$^{+0.01}_{-0.01}$ & 0.10$^{+0.08}_{-0.08}$ & 0.09$^{+0.01}_{-0.01}$ & 0.10$^{+0.04}_{-0.04}$ & 30.9$^{+4.3}_{-4.0}$ & 9.48$^{+1.11}_{-1.08}$\\ \hline 
K2-131 & 0.83$^{+0.01}_{-0.01}$ & 0.76$^{+0.01}_{-0.01}$ & -0.09$^{+0.05}_{-0.06}$ & -0.04$^{+0.02}_{-0.02}$ & -0.03$^{+0.06}_{-0.06}$ & 28.1$^{+3.2}_{-2.9}$ & 2.26$^{+1.6}_{-0.89}$\\ \hline 
K2-141 & 0.79$^{+0.01}_{-0.01}$ & 0.72$^{+0.01}_{-0.01}$ & 0.28$^{+0.09}_{-0.09}$ & 0.16$^{+0.07}_{-0.07}$ & 0.38$^{+0.01}_{-0.01}$ & 29.9$^{+4.9}_{-4.5}$ & 1.36$^{+0.63}_{-0.28}$\\ \hline 
K2-229 & 0.89$^{+0.01}_{-0.01}$ & 0.80$^{+0.01}_{-0.01}$ & 0.12$^{+0.08}_{-0.08}$ & 0.20$^{+0.02}_{-0.02}$ & 0.34$^{+0.07}_{-0.07}$ & 23.7$^{+3.8}_{-3.5}$ & 1.04$^{+0.09}_{-0.03}$\\ \hline 
K2-265 & 0.95$^{+0.01}_{-0.01}$ & 0.93$^{+0.01}_{-0.01}$ & 0.16$^{+0.07}_{-0.07}$ & 0.22$^{+0.01}_{-0.01}$ & 0.29$^{+0.06}_{-0.06}$ & 26.3$^{+3.7}_{-3.4}$ & 5.67$^{+0.8}_{-0.68}$\\ \hline 
K2-291 & 0.92$^{+0.02}_{-0.02}$ & 0.89$^{+0.01}_{-0.01}$ & 0.09$^{+0.08}_{-0.08}$ & 0.07$^{+0.04}_{-0.04}$ & 0.06$^{+0.04}_{-0.04}$ & 32.2$^{+4.3}_{-4.0}$ & 4.39$^{+1.36}_{-1.39}$\\ \hline 
K2-38 & 1.02$^{+0.01}_{-0.01}$ & 1.15$^{+0.01}_{-0.01}$ & 0.22$^{+0.07}_{-0.07}$ & 0.30$^{+0.02}_{-0.02}$ & 0.23$^{+0.06}_{-0.06}$ & 28.5$^{+3.9}_{-3.7}$ & 8.23$^{+0.35}_{-0.34}$\\ \hline 
Kepler-107 & 1.25$^{+0.01}_{-0.01}$ & 1.44$^{+0.01}_{-0.01}$ & 0.31$^{+0.11}_{-0.11}$ & 0.35$^{+0.03}_{-0.03}$ & 0.32$^{+0.08}_{-0.08}$ & 29.7$^{+6.0}_{-5.2}$ & 3.72$^{+0.36}_{-0.32}$\\ \hline 
Kepler-10 & 0.92$^{+0.02}_{-0.02}$ & 1.08$^{+0.01}_{-0.01}$ & -0.08$^{+0.06}_{-0.06}$ & 0.12$^{+0.04}_{-0.04}$ & -0.11$^{+0.05}_{-0.05}$ & 26.6$^{+3.2}_{-3.0}$ & 10.19$^{+0.95}_{-0.95}$\\ \hline 
Kepler-20 & 0.89$^{+0.02}_{-0.02}$ & 0.91$^{+0.01}_{-0.01}$ & 0.05$^{+0.08}_{-0.08}$ & 0.03$^{+0.02}_{-0.02}$ & 0.00$^{+0.03}_{-0.03}$ & 32.5$^{+4.4}_{-4.1}$ & 7.93$^{+1.48}_{-1.34}$\\ \hline 
Kepler-36 & 1.09$^{+0.02}_{-0.02}$ & 1.70$^{+0.01}_{-0.01}$ & -0.16$^{+0.06}_{-0.06}$ & -0.15$^{+0.01}_{-0.01}$ & -0.17$^{+0.07}_{-0.07}$ & 31.3$^{+3.9}_{-3.6}$ & 6.60$^{+0.34}_{-0.28}$\\ \hline 
Kepler-78 & 0.82$^{+0.0}_{-0.01}$ & 0.75$^{+0.01}_{-0.01}$ & 0.03$^{+0.07}_{-0.07}$ & 0.15$^{+0.01}_{-0.01}$ & 0.06$^{+0.05}_{-0.05}$ & 27.2$^{+3.7}_{-3.4}$ & 1.50$^{+0.65}_{-0.37}$\\ \hline 
Kepler-93 & 0.88$^{+0.02}_{-0.02}$ & 0.94$^{+0.01}_{-0.01}$ & -0.09$^{+0.06}_{-0.06}$ & 0.03$^{+0.08}_{-0.08}$ & -0.18$^{+0.04}_{-0.04}$ & 29.8$^{+3.9}_{-3.7}$ & 8.48$^{+1.21}_{-1.18}$\\ \hline 
WASP-47 & 1.03$^{+0.01}_{-0.01}$ & 1.16$^{+0.01}_{-0.01}$ & 0.41$^{+0.08}_{-0.08}$ & 0.48$^{+0.06}_{-0.06}$ & 0.45$^{+0.06}_{-0.06}$ & 27.8$^{+4.3}_{-4.0}$ & 8.58$^{+0.45}_{-0.41}$\\ \hline 
    \end{tabular}
    \caption{Host star parameters and their Fe-MF$_{st}$ equivalent value. For more details about the stellar parameters see \citetalias{paperI}.}
    \label{tab:star}
\end{table*}

\begin{table}[]
\centering 
\renewcommand{\arraystretch}{1.5} 
\begin{tabular}{|l|l|l|}
\hline
\multicolumn{1}{|c|}{With weights ($w_i$)} & \multicolumn{1}{c|}{Equal weights ($w_i=1$)} & \multicolumn{1}{c|}{No RTR planets ($\rho/\rho_\oplus>0.8$) } \\ \hline
Fe-MF$_{pl}$ = 0.97$^{+0.99}_{-1.12}$ Fe-MF$_{st}$ $-$ 0.03$^{+0.32}_{-0.28}$  & 1.83$^{+0.84}_{-1.30}$ Fe-MF$_{st}$ $-$ 0.29$^{+0.34}_{-0.22}$  & 1.29$^{+0.83}_{-0.90}$ Fe-MF$_{st}$ $-$ 0.10$^{+0.25}_{-0.23}$ \\ 
Fe-MF$_{pl}$ = 0.24$^{+0.16}_{-0.15}$ [Fe/H] $+$ 0.23$^{+0.03}_{-0.03}$  & 0.11$^{+0.14}_{-0.14}$ [Fe/H] $+$ 0.20$^{+0.03}_{-0.03}$  & 0.37$^{+0.16}_{-0.16}$ [Fe/H] $+$ 0.23$^{+0.03}_{-0.03}$ \\ 
Fe-MF$_{pl}$ = 0.13$^{+0.17}_{-0.15}$ [Mg/H] $+$ 0.23$^{+0.04}_{-0.04}$  & -0.01$^{+0.13}_{-0.12}$ [Mg/H] $+$ 0.22$^{+0.03}_{-0.03}$  & 0.20$^{+0.20}_{-0.20}$ [Mg/H] $+$ 0.23$^{+0.04}_{-0.04}$ \\ 
Fe-MF$_{pl}$ = 0.19$^{+0.13}_{-0.13}$ [Si/H] $+$ 0.23$^{+0.03}_{-0.03}$  & 0.07$^{+0.11}_{-0.11}$ [Si/H] $+$ 0.21$^{+0.03}_{-0.03}$  & 0.29$^{+0.13}_{-0.13}$ [Si/H] $+$ 0.23$^{+0.03}_{-0.03}$ \\ 
Fe-MF$_{pl}$ = -0.46$^{+0.50}_{-0.45}$ [$\alpha/Fe$] $+$ 0.24$^{+0.03}_{-0.03}$  & -0.17$^{+0.10}_{-0.10}$ [$\alpha/Fe$] $+$ 0.28$^{+0.01}_{-0.01}$  & -0.65$^{+0.45}_{-0.41}$ [$\alpha/Fe$] $+$ 0.25$^{+0.03}_{-0.03}$ \\ 
Fe-MF$_{pl}$ = 0.15$^{+0.14}_{-0.14}$ $[\alpha/H]$ $+$ 0.24$^{+0.03}_{-0.03}$  & 0.03$^{+0.10}_{-0.11}$ $[\alpha/H]$ $+$ 0.21$^{+0.02}_{-0.02}$  & 0.24$^{+0.16}_{-0.16}$ $[\alpha/H]$ $+$ 0.25$^{+0.03}_{-0.03}$ \\ 
Fe-MF$_{pl}$ = -0.02$^{+0.01}_{-0.01}$ Age $+$ 0.38$^{+0.07}_{-0.07}$  & -0.02$^{+0.01}_{-0.01}$ Age $+$ 0.38$^{+0.06}_{-0.06}$  & -0.02$^{+0.01}_{-0.01}$ Age $+$ 0.40$^{+0.07}_{-0.07}$ \\ 
\hline \hline
$\rho/\rho_\oplus = $ 0.97$^{+1.05}_{-1.12}$ Fe-MF$_{st}$ $+$ 0.65$^{+0.32}_{-0.30}$  & 2.59$^{+0.83}_{-0.90}$ Fe-MF$_{st}$ $+$ 0.17$^{+0.24}_{-0.22}$  & 1.25$^{+0.85}_{-0.88}$ Fe-MF$_{st}$ $+$ 0.60$^{+0.25}_{-0.24}$ \\ 
$\rho/\rho_\oplus = $ 0.21$^{+0.15}_{-0.14}$ [Fe/H] $+$ 0.91$^{+0.03}_{-0.03}$  & -0.26$^{+0.09}_{-0.09}$ [Fe/H] $+$ 0.90$^{+0.03}_{-0.03}$  & 0.32$^{+0.17}_{-0.16}$ [Fe/H] $+$ 0.93$^{+0.03}_{-0.03}$ \\ 
$\rho/\rho_\oplus = $ 0.13$^{+0.16}_{-0.14}$ [Mg/H] $+$ 0.91$^{+0.04}_{-0.04}$  & -0.26$^{+0.08}_{-0.08}$ [Mg/H] $+$ 0.92$^{+0.03}_{-0.03}$  & 0.22$^{+0.20}_{-0.19}$ [Mg/H] $+$ 0.92$^{+0.04}_{-0.04}$ \\ 
$\rho/\rho_\oplus = $ 0.15$^{+0.12}_{-0.12}$ [Si/H] $+$ 0.92$^{+0.03}_{-0.03}$  & -0.22$^{+0.08}_{-0.08}$ [Si/H] $+$ 0.90$^{+0.03}_{-0.03}$  & 0.24$^{+0.14}_{-0.13}$ [Si/H] $+$ 0.93$^{+0.03}_{-0.03}$ \\ 
$\rho/\rho_\oplus = $ -0.45$^{+0.50}_{-0.49}$ [$\alpha/Fe$] $+$ 0.92$^{+0.03}_{-0.03}$  & -1.25$^{+0.30}_{-0.37}$ [$\alpha/Fe$] $+$ 0.87$^{+0.03}_{-0.03}$  & -0.54$^{+0.39}_{-0.42}$ [$\alpha/Fe$] $+$ 0.94$^{+0.03}_{-0.03}$ \\ 
$\rho/\rho_\oplus = $ -0.04$^{+0.15}_{-0.15}$ [$\alpha$/H] + 0.94$^{+0.03}_{-0.03}$ & 0.39$^{+0.05}_{-0.05}$ [$\alpha$/H] + 0.85$^{+0.01}_{-0.01}$  &  -0.35$^{+0.06}_{-0.07}$ [$\alpha$/H] + 1.02$^{+0.02}_{-0.02}$ \\ 
$\rho/\rho_\oplus = $ -0.02$^{+0.01}_{-0.01}$ Age $+$ 1.04$^{+0.07}_{-0.06}$  & -0.03$^{+0.01}_{-0.01}$ Age $+$ 1.09$^{+0.06}_{-0.05}$  & -0.02$^{+0.01}_{-0.01}$ Age $+$ 1.06$^{+0.07}_{-0.06}$ \\ 
\hline 

\end{tabular}
\caption{Table of linear fit coefficients for this data considering: with weights ($w_i$), without weights ($w_i=1$) and without exoplanet above RTR ($\rho/\rho_\oplus>0.8$). {In comparison \citetalias{Adibekyan2021} and \citetalias{Brinkman2024} find Fe-MF$_{pl}$ = 6.8$^{+1.3}_{-1.3}$ Fe-MF$_{st}$ $-$ 1.84$^{+0.4}_{-0.4}$ and Fe-MF$_{pl}$ = 5.6$^{+1.6}_{-1.6}$ Fe-MF$_{st}$ $-$ 1.4$^{+0.6}_{-0.6}$, respectively using ODR linear fitting method.}}
\label{tab:coef_ext}
\end{table}

\end{document}